\documentclass[structabstract]{aa}
\usepackage{graphicx}
\usepackage{txfonts}
\usepackage{lscape}
\usepackage{natbib}
\usepackage{color}
\bibpunct{(}{)}{;}{a}{}{,} 
\def\tablefoot#1{\par\vspace*{2ex}%
 \parbox{\hsize}{\leftskip0pt\rightskip0pt
 {\noindent\small\textbf{Notes.}~#1\par}}}
\def\tablefoottext#1#2{$^{(#1)}$~#2}
\def\tablefootmark#1{$^{#1}$\,\ignorespaces}

\begin{document}
   \title{Abundances of PAHs in the ISM: 
Confronting Observations with Experimental Results
\footnote{Based on observations collected at the European Southern
Observatory, Chile (079.C-0597(A))}
}

   \author{R. Gredel
          \inst{1}
          \and
          Y. Carpentier \inst{2}
          \and
          G. Rouill\'e \inst{2}
          \and
          M. Steglich \inst{2}
          \and
          F. Huisken \inst{2}
          \and
          Th. Henning \inst{1}
          }

   \institute{Max Planck Institute for Astronomy (MPIA),
              K\"onigstuhl 17, D-69117 Heidelberg\\
         \and
             Laboratory Astrophysics Group of the Max Planck Institute for Astronomy at the Friedrich Schiller
University Jena, Institute of Solid State Physics, Helmholtzweg 3, D-07743 Jena\\
             }

   \date{Received; accepted }


  \abstract
   {
The identification of the carriers of the diffuse interstellar
bands (DIBs) is the longest standing problem in the study of the
interstellar medium. {Here we present
recent UV laboratory spectra of various polycyclic aromatic
hydrocarbons (PAHs) and explore the potential of these molecules
as carriers of the DIBs. Whereas, in the near IR range, the PAHs
exhibit vibrational bands which are not molecule-specific, their
electronic transitions occurring in the UV/vis provide
characteristic fingerprints. The comparison of laboratory spectra
calibrated in intensity with high S/N observational data in the UV
enables us to establish new constraints on PAH abundances.}}
   {
From a detailed comparison of gas-phase and Ne-matrix absorption
spectra of anthracene, phenanthrene, pyrene, 2,3-benzofluorene,
benzo[ghi]perylene, and hexabenzocoronene with new interstellar
spectra, we aim to infer the abundance of these PAHs in the
interstellar medium. }
   {
We present spectra of PAHs
measured at low temperature in the gas phase and in a Ne matrix, and present methods
to derive absolute absorption cross sections for the matrix and gas phase spectra.
We have obtained high signal to noise (S/N $>$ 100) absorption spectra
toward five lines of sight with reddenings of $E_\mathrm{B-V}$ = 1 - 1.6 mag.
The spectra cover the 3050 - 3850 $\AA$ wavelength region
where the PAHs studied here show prominent absorption features.
    }
{From the observations, we infer upper limits in the fractional
abundances of the PAHs studied here.Upper limits in the column
densities of anthracene of $0.8 - 2.8 \times 10^{12}$\ cm$^{-2}$
and of pyrene and 2,3-benzofluorene ranging from $2 - 8 \times
10^{12}$\ cm$^{-2}$ are inferred. Upper limits in the column
densities of benzo[ghi]perylene are $0.9 - 2.4 \times 10^{13}$ and
$10^{14}$ cm$^{-2}$ for phenanthrene.  The measurements indicate
fractional abundances of anthracene, pyrene, and 2,3-benzofluorene
of a few times $10^{-10}$. Upper limits in the fractional
abundance of benzo[ghi]perylene of a few times $10^{-9}$ and of
phenanthrene of few times $10^{-8}$ are inferred.
{Toward CPD $-32^\circ 1734$, we found near 3584
{\AA} an absorption line of OH$^+$, which was discovered in the
interstellar medium only very recently.}}
   {The fractional abundances of PAHs inferred here are up to two orders of magnitude lower
than estimated total PAH abundances in the interstellar medium. This indicates that either
neutral PAHs are not abundant in translucent molecular clouds, or that a PAH population
with a large variety of molecules is present.}


   \maketitle
%

\section{Introduction}

The long-standing mystery concerning the nature of the carriers of
the diffuse interstellar bands (DIBs) has received considerable
and renewed attention in recent years after laboratory spectra of
potential candidates have become available. Most scientists in the
field support the general idea that the DIBs arise from electronic
transitions of gas-phase molecules. Recent examples include the
$2^2_0$, $2^1_04^1_0$, and $2^1_0$ transitions in the
$B^1\mathrm{B}_1 \leftarrow X^1\mathrm{A}_1$ system of
$l$-C$_3$H$_2$ \citep{maier10}, the (0,0) band of the $A^2\Pi_u -
X^2\Pi_g$ system of the HC$_4$H$^+$ diacetylene cation
\citep{krelowski10a}, or the  anthracene and naphtalene cations
\citep{IglesiasGroth08,IglesiasGroth10}. Similarly, a recent
theoretical work established an excellent agreement between the
$0^0_0$ band of the $^1B_1 - X^1A'$ electronic system of
CH$_2$CN$^-$ with the 8037 $\AA$ DIB \citep{cordiner07}.

DIBs are generally observed in absorption against the continua of reddened
supergiants in the Galaxy and in extragalactic sources.
The appearance of DIBs in translucent molecular clouds
characterized by a wide range of physical parameters
indicates that DIBs arise from chemically stable carriers.
More than 400 DIBs have been identified toward a single
galactic line of sight \citep{hobbs09}. The individual
absorption profiles and line strength ratios of DIBs may vary significantly from one
sightline to another \citep{galazutdinov08}.
While the equivalent widths of several DIBs correlate well with
the reddenings $E_\mathrm{B-V}$ of the background stars, the majority of the DIB
carriers appear to be enhanced in the outer regions of translucent clouds
\citep{cami97,sonnentrucker97}. Observations of DIBs in
the Magellanic Clouds \citep{welty06} have shown that some DIBs
are significantly weaker than toward Galactic lines of sight with similar reddening,
suggesting that the abundance of the responsible carriers scales with the
metallicity of the environment. General scaling relations
have not been established, however, as the equivalent widths of
the family of C$_2$-DIBs, for example, are as strong
as in the Galaxy, for lines of sight with comparable reddening \citep{welty06}.

The vibronic progressions established for several DIBs
\citep{duley10} point toward an origin from a torsional motion of
pendent rings, suggesting that the carriers are 'floppy'
polycyclic aromatic hydrocarbon (PAH) structures such as
9-phenylanthracene, parasubstituted biphenyls, tetracene
derivatives, bianthracene, tolane, etc., rather than compact
molecules. The presence of PAHs in the interstellar medium is now
generally established, and there is unequivocal evidence that the
unidentified near-infrared emission bands arise from PAHs
{\citep{Leger84,Allamandola85,Tielens08,Bauschlicher10}}.
Mixtures of PAHs or nano-sized hydrogenated dust particles have
been proposed to explain the 2175 $\AA$ UV bump seen toward stars
reddened by translucent material
\citep{joblin92,mennella98,Steglich2010}, and the presence of
carbonaceous dust particles in translucent material is well
established \citep{henning98}. The first detection of the C$_{60}$
and C$_{70}$ fullerenes in the young planetary nebula Tc~1 has
recently been reported by \citet{cami10}.

{With the present study, we have undertaken a new
effort to search for PAHs in translucent molecular clouds. While
ionized PAHs exhibit absorption features in the visible and near
IR, making them attractive carrier candidates for the DIBs
\citep{Salama99}, the electronic transitions of their neutral
counterparts are shifted to the UV. Since neutral PAHs can be
prepared more easily with sufficient density in the gas phase and
because observational spectra in the UV can be obtained with high
S/N, we focused our interest on this class of molecules. In this
paper, we present results on our} search for anthracene,
phenanthrene, pyrene, 2,3-benzofluorene, and benzo[ghi]perylene.
These molecules were measured in the gas phase with our supersonic
jet/CRDS apparatus and show prominent absorption features at
wavelengths shorter than
 4000 $\AA$. We have obtained high signal-to-noise ratio (S/N $>$ 100) spectra
toward heavily reddened supergiants in the 3050 -- 3850 $\AA$
region using UVES.  We developed techniques to calibrate the CRDS
spectra against matrix and solution spectra and to derive absolute
absorption cross sections for vibronic bands measured in the gas
phase. We also searched for hexabenzocoronene, for which we only
have Ne {and Ar} matrix spectra. We present a
method to estimate gas phase transition wavelengths from the
matrix spectra. These methods are described in detail in
Sect.~\ref{method}, which also provides the measured gas phase
absorption wavelengths and the estimated absorption cross sections
of the PAHs studied here. The astronomical spectra, obtained
toward early-type supergiants with reddenings up to
$E_\mathrm{B-V} = 1.6$ mag, are presented in
Sect.~\ref{observations}. A detailed comparison with the
laboratory results is given in Sect.~\ref{comparisons}, and
Sect.~\ref{sectionupperlimits} presents upper limits of PAH column
densities toward the lines of sight studied here.
Section~\ref{islines} contains an analysis of various diatomic
molecular absorption lines which are detected toward the observed
stars, and which are used for consistency checks with previous
results. The inferred upper limits and the corresponding maximal
fractional abundances of anthracene, phenanthrene, pyrene,
2,3-benzofluorene, benzo[ghi]perylene, and hexabenzocoronene are
compared to expectations of PAH abundances in 
translucent material in Sect.~\ref{pahabundance}.

\section{Laboratory Data}
\label{method}

\renewcommand{\thefootnote}{\alph{footnote}} 

The confirmation of potential DIB carrier molecules requires
laboratory spectra that are obtained under conditions that
match as close as possible the conditions in interstellar clouds.
Such conditions - collision-free
environments and low temperatures - prevail in the expansion of
supersonic jets of rare gases seeded with the molecule to be
studied. In combination with cavity ring-down laser absorption spectroscopy
(CRDS) {\citep{OKeefe88}}, low-temperature gas phase absorption spectra are obtained,
providing the ultimate data which can be compared with
astrophysical observations.
For several reasons, however, supersonic jet CRDS spectra are
difficult to obtain. It is for instance difficult to transfer
complex molecules into the gas phase. In such cases, another
low-temperature technique, matrix isolation spectroscopy (MIS)
{\citep{Bondybey96}}, is performed. In MIS
experiments, the sample molecules are dispersed in a transparent
matrix of rare gas atoms such as Ne or Ar kept at cryogenic
temperatures of 6.5~K (Ne-matrix) and 12~K (Ar-matrix).  Compared
to laser spectroscopy in supersonic jets, the MIS technique has
the advantage that wide ranges of wavelengths are scanned at once
and that only small amounts of the sample molecules are required
in general. The main disadvantage of MIS experiments is that the
molecules are no longer in a collision-free environment.
Interactions with the surrounding matrix material induce shifts of
the transition energies and a broadening of the absorption
profiles. Spectra obtained from MIS experiments are thus of
limited use in the astrophysical context. On the other hand, CRDS
experiments do not generally allow to obtain absorption cross
sections, as it is difficult to measure the density of the
relevant molecules in the supersonic jets. This is more easily
achieved for MIS spectra which can be readily compared with
solvent spectra, where the concentration of the dissolved
molecules can be accurately {measured. We have
developed methods to derive} absolute absorption cross sections
for supersonic jet spectra and to determine gas phase absorption
wavelengths from MIS spectra, and present them in
Sects.~\ref{lab-sigma} and \ref{lab-wavelengths} below.

\begin{figure}
  \includegraphics[width=0.5\linewidth]{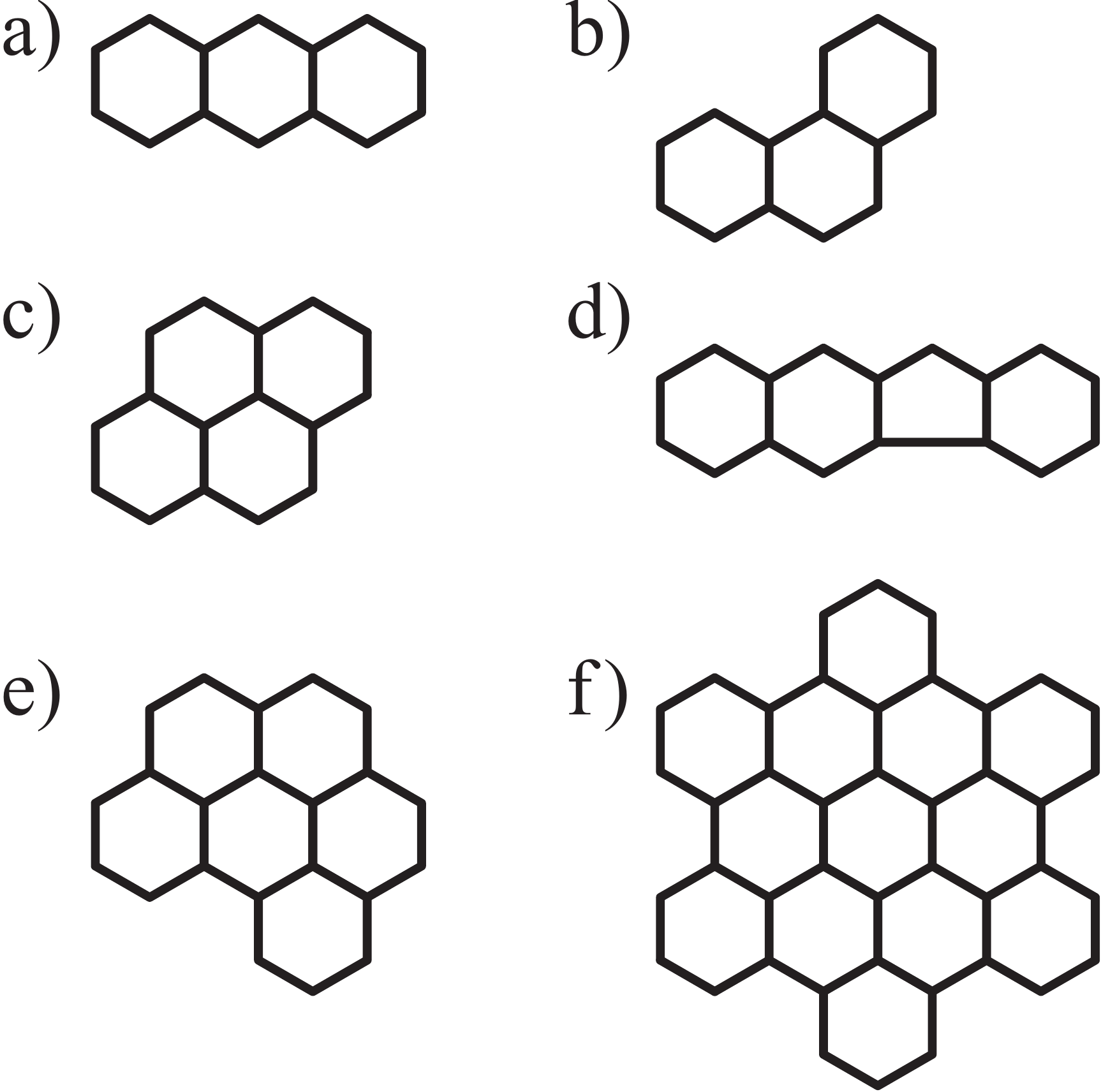}
  \caption{Schematic carbon structures of the PAHs discussed in this study. \textbf{a)}: anthracene,
C$_{14}$H$_{10}$; \textbf{b)}: phenanthrene, C$_{14}$H$_{10}$; \textbf{c)}: pyrene, C$_{16}$H$_{10}$; \textbf{d)}:
2,3-benzofluorene, C$_{17}$H$_{12}$; \textbf{e)}: benzo[ghi]perylene, C$_{22}$H$_{12}$; \textbf{f)}:
hexabenzocoronene, C$_{42}$H$_{18}$.}
  \label{PAHstructures}
\end{figure}

From the various PAHs studied in our laboratory, we have selected
six species revealing prominent absorption features in the 3050 -
3850 $\AA$ region, and for which CRDS spectra are available
{\citep{Rouille04, Rouille07, Staicu04, Staicu08}}.
In the order of increasing size, these molecules are anthracene,
phenanthrene, pyrene, 2,3-benzofluorene, benzo[ghi]perylene, and
hexa-{\it peri}-hexa\-benzocoronene (hexabenzocoronene in the
following). The structures of the six molecules are illustrated in
Fig.~\ref{PAHstructures}. Calibrated solution spectra obtained in
the nonpolar solvent cyclohexane for anthracene, phenanthrene,
pyrene, and benzo[ghi]perylene \citep{Karcher1985} and in
1,2,4-trichlorobenzene for hexabenzocoronene \citep{Hendel1986}
are used {as references for the determination of
absolute absorption cross sections as described in the next
section. The solution spectrum of 2,3-benzofluorene has been
measured in our laboratory, using cyclohexane as solvent. Spectra
of anthracene and pyrene have been measured as well and compared
with the literature data to validate our experimental setup.} The
matrix isolation spectra of the six molecules were recorded in Ne
at 6.5~K with an apparatus described elsewhere
\citep{Steglich2010}. For hexabenzocoronene, the calibration
procedure has been restricted to a Ne matrix spectrum
\citep{Rouille2009} since a jet spectrum in the wavelength range
of interest is not available.

\subsection {Determination of absolute absorption cross sections
for supersonic jet spectra}
\label{lab-sigma}

In cavity ring-down spectroscopy (CRDS) experiments, the
determination of vibrational absorption cross sections is
generally not possible, as the determination of the densities in
the supersonic jets remains difficult. In contrast, absolute
absorption cross sections are routinely obtained from solvent
spectra \citep{Karcher1985}, where the absorption is expressed in
terms of a molar extinction coefficient $\epsilon$ (in $\rm
l~mol^{-1}~cm^{-1}$). {Indeed, assuming that
absorption is the only significant contribution to the extinction,
$\epsilon$ can be converted into an} absorption cross section
$\sigma$ (in cm$^2$) according to
 \begin{equation}
\sigma = \frac{1000 \ln{10}}{N_A} \epsilon = 3.8235 \times 10^{-21} \epsilon
\end{equation}
where $N_A$ is the Avogadro constant. {By
integrating $\sigma$ over a wavelength interval that includes a
specific absorption band, one obtains the integrated cross section
of the band. It can} be used to derive molecular column densities
from interstellar spectra using the formalism summarized in
Sect.~\ref{sectionupperlimits} (in particular Eq.~\ref{eqno2}).

\begin{figure}
    \begin{center}
        \includegraphics[width=0.9\linewidth]{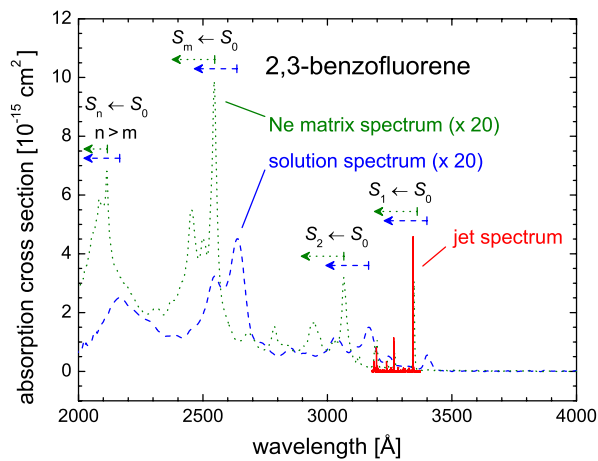}
    \end{center}
    \caption{Absorption spectra of 2,3-benzofluorene measured in a supersonic jet (red curve), in a Ne matrix at
6.5 K (green curve), and in cyclohexane at room temperature (blue curve). Note that the matrix and solution spectra
were magnified by a factor of 20. All spectra were measured in our laboratory.}
    \label{fig:BZF_wavelength}
\end{figure}

\begin{figure}
    \begin{center}
        \includegraphics[width=0.9\linewidth]{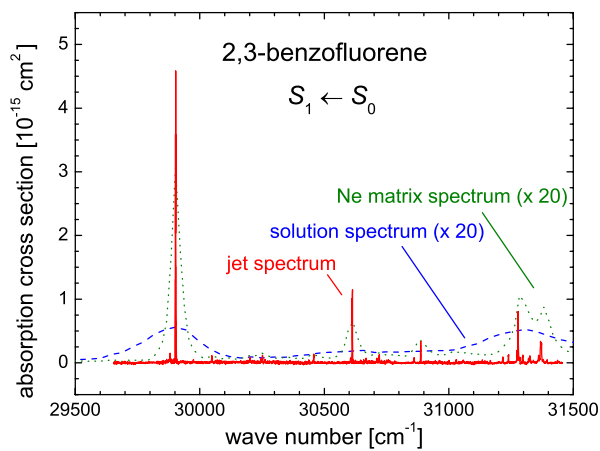}
    \end{center}
    \caption{$S_1 \leftarrow S_0$ absorption spectra of 2,3-benzofluorene under the same experimental conditions
as in Fig.~\ref{fig:BZF_wavelength}. The solution and matrix spectra have been shifted to higher energy by
491~cm$^{-1}$ and 34~cm$^{-1}$, respectively.}
    \label{fig:BZF_wavenumber}
\end{figure}

Solvents modify the band intensities of the dissolved molecules
and cause shifts of the transition wavelengths and a broadening of
the vibronic bands. The effect of the wavelength shifts will be
addressed in detail in Sect.~\ref{lab-wavelengths}. Due to the
broadening, an absorption peak measured in solution may correspond
to several vibronic bands. However, if solution spectra are
sufficiently well-resolved, they may be used to derive estimates
of the vibrational absorption cross section for matrix spectra
\citep{Ehrenfreund1992} or directly for jet spectra
\citep{Kokkin2008}.  The method is based on the
{reasonable assumption that the absorption cross
section of a molecule integrated over an entire electronic
transition is independent of the environment of the molecule, i.e.
whether it is vacuum, a rare gas matrix, or a solution. The error
that we introduce when we make this assumption is discussed in the
next paragraph. For the calibration of our high-resolution gas
phase spectra, we have developed a procedure where we use a matrix
spectrum as an intermediate step between the solution spectrum and
the jet spectrum.} We integrate the complete region corresponding
to an electronic transition in the solution and matrix spectra and
equal them. The same operation is then repeated for the single
band of interest in the matrix and jet spectra. The calibration
procedure is exemplified in Figs.~\ref{fig:BZF_wavelength} and
\ref{fig:BZF_wavenumber} for the case of 2,3-benzofluorene. Both
figures show the measured absorption profiles of 2,3-benzofluorene
in a Ne matrix, in a solution, and in a jet.  The different shifts
of the band groups seen in Fig.~\ref{fig:BZF_wavelength} allow us
to distinguish between the different electronic transitions of the
molecule. The calibration has been applied exclusively to the
first electronic transition ($S_1 \leftarrow S_0$) of
2,3-benzofluorene. The direct comparison of the areas of the
{higher-energy transitions} in the solution and
matrix spectra is actually complicated by the presence of a
continuum in the matrix spectrum attributed to the scattering by
the PAH molecules themselves. In Fig.~\ref{fig:BZF_wavenumber},
the spectra of the first electronic transition are displayed in
wave number, and the matrix and solution spectra are shifted to
the blue to account for environment effects (34 and 491 cm$^{-1}$,
respectively). To calibrate the matrix spectrum against the
solution spectrum, the integration has been performed from 29~500
to 31~500 cm$^{-1}$ (see Fig.~\ref{fig:BZF_wavenumber}). The area
of the gas phase origin band at 29~894 cm$^{-1}$ has then been
equaled to the corresponding band at 3348~\AA{} in the matrix
spectrum while the contribution of the smaller band at 29~872
cm$^{-1}$ has been subtracted.

{In order to estimate the error that we introduce
with the assumption that the {\it integrated} cross section is not
affected by the environment of the molecule, we have to consider
the electronic interaction between the solute molecule and the
neighboring solvent molecules. The variations of the band
intensities caused by this interaction are described by various
models \citep{Chako1934,Linder1971}.  In the first approximation,
the Lorentz-Lorenz force which arises from the polarization of the
surrounding solvent molecules can be described in terms of the
Chako factor $9n / \left( n^2 + 2 \right)^2$, where $n$ is the
refractive index of the solvent. Multiplying the oscillator
strength of a band measured in solution by the Chako factor, the
corresponding oscillator strength in the gas phase is obtained.
These models have been successfully applied to C$_{60}$ by
\citet{Smith96}. For the wavelength range and the experiments
discussed here, where cyclohexane was used as solvent, the Chako
factor is 0.78. In view of the fact that the Chako factor is still
an approximation and that the proper ratio of the oscillator
strengths in the gas phase and in a solution depends on the exact
nature of both the solvent and solute molecules, we chose not to
apply any correction factor to our cross section values. Taking
into account the other sources of errors including the
experimental and data processing uncertainties, we estimate that
the oscillator strengths are accurate within $\pm 50$ \%.}

\begin{table*}
    \caption{Spectral properties (attribution, position, width, integrated cross section, and oscillator
strength) of the PAH bands considered for comparison with astrophysical data in this article.}
        \begin{tabular}{lcccccc}
        \hline
        \hline
        Molecule           & Electronic transition & $\lambda_{\rm{air}}$\tablefootmark{a} & FWHM    & $\int \sigma d\lambda$     & $f$ & $N_{\rm{max}}$\tablefootmark{b} \\
                           &                       & ({\AA})                               & ({\AA}) & ($10^{-16}$ cm$^2$ {\AA})  &     & ($\textrm{cm}^{-2}$) \\
        \hline
        anthracene         & $S_1 \leftarrow S_0$  & 3610.74 \tablefootmark{c} & 0.34   & 19     & 0.016   & $5.3 \times 10^{11}$ \\
        phenanthrene       & $S_1 \leftarrow S_0$  & 3409.21 \tablefootmark{d} & 0.20   & 0.16   & 0.00016 & $6.2 \times 10^{13}$ \\
        pyrene             & $S_2 \leftarrow S_0$  & 3208.22 \tablefootmark{e} & 5.5    & 89     & 0.097   & $1.1 \times 10^{11}$ \\
                           &                       & 3166.26 \tablefootmark{e} & 4.7    & 30     & 0.033   & $3.4 \times 10^{11}$ \\
        2,3-benzofluorene  & $S_1 \leftarrow S_0$  & 3344.16 \tablefootmark{f} & 0.18   & 8.4    & 0.0085  & $1.2 \times 10^{12}$\\
                           &                       & 3266.63 \tablefootmark{f} & 0.22   & 2.4    & 0.0025  & $4.2 \times 10^{12}$\\
        benzo[ghi]perylene
                           & $S_2 \leftarrow S_0$  & 3512.15 \tablefootmark{g} & 8.2    & 9.9    & 0.009   & $1 \times 10^{12}$\\
                           &                       & 3501.76 \tablefootmark{g} & 5.3    & 5.7    & 0.005   & $1.8 \times 10^{12}$\\
        hexabenzocoronene  & $\beta$-band          & $3344\pm8$ \tablefootmark{h}   &        & 900 \tablefootmark{i} &         & \\
        \hline
        \end{tabular}
\tablefoot{
\tablefoottext{a}{Band positions are
given for air under standard conditions (see text).}
\tablefoottext{b}{An upper limit of the column density has been
calculated assuming a detection limit with an equivalent width of
1 m{\AA}.} \tablefoottext{c}{Staicu et al. (2004).}
\tablefoottext{d}{Rouill\'e et al. (2010).}
\tablefoottext{e}{Rouill\'e et al. (2004).}
\tablefoottext{f}{Staicu et al. (2008).}
\tablefoottext{g}{Rouill\'e et al. (2007).}
\tablefoottext{h}{Extrapolated from Ne and Ar matrix measurements
(Rouill\'e et al. 2009b).} \tablefoottext{i}{Calculated for Ne
matrix spectra.} }
    \label{tab:cross-section}
\end{table*}

The results of the calibration procedure are summarized in
Table~\ref{tab:cross-section}. The wavelength positions and widths
are from our CRDS experiments, except for hexabenzocoronene, for
which the wavelength position in the gas phase was extrapolated
from matrix measurements (cf. Sect.~\ref{lab-wavelengths}).
{Although the band positions were originally
determined for vacuum, we converted them to air because
high-resolution astronomical spectra are traditionally calibrated
against Th/Ar lines whose wavelengths are given for air. The
conversion was carried out according to $\lambda_{\mathrm{air}} =
\lambda_{\mathrm{vac}}/n$ with $n$ being the index of refraction
of air under standard conditions at $\lambda_{\mathrm{vac}}$, as
computed with Eq. A1 of \citet{Livengood99}.} For anthracene and
phenanthrene, the results are given for the origin of their
respective $S_1 \leftarrow S_0$ transitions. For pyrene, the
spectrum of the $S_2 \leftarrow S_0$ transition is composed of
band clusters due to the vibronic interaction between the $S_2$
and $S_1$ states. We give in Table~\ref{tab:cross-section} the
central wavelengths and the total FWHMs of the two band groups at
3166 $\AA$ and 3208 $\AA$ \citep{Rouille04}. For
2,3-benzofluorene, the origin and a strong vibronic band of the
$S_1 \leftarrow S_0$ transition are listed. For
benzo[ghi]perylene, the spectral region around 3500~\AA{} shows
two broad bands attributed to vibrationally excited bands of the
$S_2  \leftarrow S_0$ transition. We also list the oscillator
strength $f$ obtained in the approximation of weak bands
\citep{Mulliken} from
\begin{equation}
    f = \frac{4 \epsilon_0 m c^2}{e^2 \lambda^2} \int{\sigma(\lambda) d\lambda},
\end{equation}
where $\lambda$ is the central wavelength of the band and
$\epsilon_0$, $c$, $m$, and $e$ are the vacuum permittivity, the
speed of light in vacuum, the mass, and the charge of the
electron, respectively. {It is interesting to note
that, in some cases, the absorption cross sections and oscillator
strengths vary significantly for the different molecules studied
especially if we consider the transition with the lowest energy
($S_1 \leftarrow S_0$). Thus, we found that the oscillator
strengths of this transition differ by two orders of magnitude if
we compare the two molecules anthracene and phenanthrene, which
are both composed of three aromatic cycles (see Figure 1). This
shows that the change of the electronic structure during the
transition, which determines the strength of the transition, does
not obey simple empirical rules. In fact, laboratory experiments
seem necessary to determine this important property.}

In the case of 2,3-benzofluorene, an oscillator strength of 0.0085
is derived for the origin band of the $S_1 \leftarrow S_0$
transition (see Table~\ref{tab:cross-section}). The summation over
the complete electronic transition (\emph{i.e.} the origin band
and the associated vibronic bands) yields an
{integrated oscillator strength of 0.03. This value
is compared in Table~\ref{tab:calculations} with the results of
theoretical calculations \citep{Staicu08} and a laboratory
measurement \citep{Banisaukas04}. Theoretical transition energies
and oscillator strengths were obtained by applying different
quantum chemistry models. In Table~\ref{tab:calculations}, we
refer to results obtained with a semiempirical model, ZINDO, an
\textit {ab initio} method, CIS(D)/6-31G(d), and a model based on
the time-dependent density functional theory, TD-DFT-B3LYP/TZ. For
details on the computational methods, see \citet{Staicu08} and
references therein.} As far as the oscillator strength is
concerned, the largest deviation from experimental data is found
for the ZINDO calculation although it yields the best result for
the transition energy. The oscillator strengths of the other two
calculations come close to the experimental values. The
satisfactory agreement of our oscillator strength with \textit {ab
initio} calculations and the result derived from Ar matrix data
gives us confidence that our method to determine absolute
absorption cross sections leads to reasonable results.

For the other PAHs discussed in this paper, absorption cross
sections of calculated spectra are available in a database
\citep{Malloci04, Malloci07}. {Unfortunately, the
theoretical spectra were convoluted with a too broad line shape
function making a comparison with our experimental data very
difficult or even impossible.}

{At this point, a general statement about computed
electronic transition energies seems appropriate. As can be
inferred from the data of Table~\ref{tab:calculations}, the
theoretical values may strongly differ from the experimental
results. The current quantum chemistry models and calculation
techniques allow one to predict electronic transition energies
with an accuracy of the order of 0.2 eV at best. The accuracy of
semiempirical calculations, even if the models are well
parametrized, suffers from their simplicity. On the other hand,
the {\it ab initio} approach is the most complete. In order to be
applied to large molecules, however, approximations have to be
made, affecting the accuracy of the results. Finally, calculations
based on the density functional theory currently miss functionals
adapted to PAH molecules. For our example molecule,
2,3-benzofluorene, the best result is obtained with the
semiempirical ZINDO method. However, the deviation from the
experimental value is with more than 20 {\AA} still quite
significant. In comparison, the extrapolation of matrix data, an
experimental method which will be described in the next section,
yields a frequency position as close as 5 {\AA} to the gas phase
position. In conclusion, we can say that the uncertainty of
quantum chemical calculations is still too large for a meaningful
comparison with high-precision astronomical observations.}

\begin{table}
    \caption{Comparison of the energies and oscillator strengths for the first
electronic transition of 2,3-benzofluorene in calculated and experimental
spectra.}
        \begin{tabular}{lccc}
        \hline
        \hline
        Chemistry model & Energy    & $f$    & Ref. \\
                        & cm$^{-1}$ &        & \\
        \hline
        ZINDO           & 30095     & 0.0116 & $^a$ \\
        CIS(D)/6-31G(d) & 34923     & 0.0234 & $^a$ \\
        TD-DFT-B3LYP/TZ & 30431     & 0.038  & $^a$ \\
        Ar matrix       & 29647     & 0.024  & $^b$ \\
        supersonic jet  & 29894.3   & 0.03   & $^c$ \\
        \hline
        \multicolumn{3}{l}{$^a$ \cite{Staicu08}}\\
        \multicolumn{3}{l}{$^b$ \cite{Banisaukas04}}\\
        \multicolumn{3}{l}{$^c$ \cite{Staicu08} \& this work}\\
        \end{tabular}
    \label{tab:calculations}
\end{table}

\subsection {Determination of gas phase absorption wavelengths from MIS spectra}
\label{lab-wavelengths}

PAH absorption wavelengths obtained from MIS experiments or from
solutions suffer from a broadening and a shift of the absorption
features, compared to the absorption profiles obtained in an
environment free of interactions. {It is
nevertheless possible to exploit MIS results to predict band
positions in the gas phase if they are not available, as in the
case of the $\beta$-band of hexabenzocoronene
\citep{Rouille2009}.} For PAH molecules in rare gas matrices, the
dominating mechanism in the interaction between the molecules and
the rare gas atoms is the dispersion effect. The effect arises
from the interaction of polarizable species and is proportional to
the product of the mean static dipole polarizabilities
\citep{London1937}. As the dispersion effect depends on the
polarizability of the matrix material, a given transition shows
different energy shifts in different matrices, and an
extrapolation to a matrix with zero polarizability allows one to
estimate the band position in the gas phase. The models of
{\citet{Longuet-Higgins1957} and \citet{Shalev91a,
Shalev91b, Shalev92}} have been applied to evaluate the transition
energy shifts for PAH molecules surrounded by rare gas atoms
\citep{Biktchantaev2002, Shalev91a, Shalev91b, Shalev92}.
{These models take into account the arrangement of
the rare gas atoms around the molecules. As this information
cannot be obtained by straightforward methods, we assume that the
dispersion effect does not vary significantly when we consider the
different geometries encountered for PAH molecules in Ne and Ar
matrices.} The transition frequency
$\tilde{\nu}_\mathrm{gas\:phase}$ of a band is then obtained as
\begin{equation}
\tilde{\nu}_{\mathrm{gas\:phase}} = \tilde{\nu}_{\mathrm{Ne\:matrix}} +  \frac{1}{R_\alpha-1} \left( \tilde{\nu}_{\mathrm{Ne\:matrix}} - \tilde{\nu}_{\mathrm{Ar\:matrix}} \right) ,
\end{equation}
where $R_\alpha = 4.13$ is
the ratio of the polarizabilities of Ar and Ne \citep{Radzig85}.

We apply this method to the complete electronic spectrum of
hexabenzocoronene that we have recently measured in a Ne matrix
\citep{Rouille2009}. The full spectrum reveals strong absorption
bands in the $3000-4000$~{\AA} spectral region. A well defined
feature of the $\alpha$-band was found at 4383 $\pm$ 5 and 4344
$\pm$ 5 {\AA} in Ar and Ne, respectively. The linear extrapolation
based on the ratio of the rare gas polarizabilities leads to a
wavelength of 4332 $\pm$ 9 {\AA}. This feature was measured in a
cold molecular beam \citep{Kokkin2008} at a vacuum wavelength of
4335.2 {\AA}, which corresponds to 4334.0 {\AA} in air, a value in
close agreement with our prediction. The significantly stronger
$\beta$ band of hexabenzocoronene occurs near 3344 $\AA$
\citep{Rouille2009}.

Unlike the $\alpha$-band, the $\beta$-band shows features that are
unresolved in the matrix spectra. They correspond to numerous
bands that arise through the interaction of the $S_\beta$
electronic state with the vibrational manifold of $S_2$
\citep{Rouille2009}. As a consequence, each peak of the
$\beta$-band in the matrix spectra is expected to be resolved in
the gas phase into a group of bands. For example, such groups are
found in the spectrum of the $S_2 \leftarrow S_0$ transition of
pyrene (cf. Sect.~\ref{comparisons}). In the case of pyrene, the
extrapolation from peaks {measured by us at 3302
$\pm$ 4 and 3234 $\pm$ 5 $\AA$ in Ar and Ne matrices,
respectively,} gives an expected position of 3213 $\pm$ 8 {\AA} of
this band in the gas phase. The strongest peak of this group has
been measured at 3208.2 {\AA} or 31~160.9 $\pm$ 0.5 cm$^{-1}$ in
the gas phase \citep{Rouille04}, in good agreement with the result
of the extrapolation. Considering now the matrix spectra of
hexabenzocoronene, the relative intensities of the peaks in the
$\beta$-band vary depending on the matrix material. We assume that
the strongest peak in the Ne matrix spectrum corresponds in the
gas phase to the group of bands that includes the strongest one.
The peak lies at 3420 $\pm$ 4 and 3362 $\pm$ 5 {\AA} in Ar and Ne
\citep{Rouille2009}, respectively. Our simple model leads to an
estimated wavelength of 3344 $\pm$ 8 {\AA} of this band in the gas
phase.

   \begin{figure}
   \centering
   \includegraphics[angle=0,width=10cm]{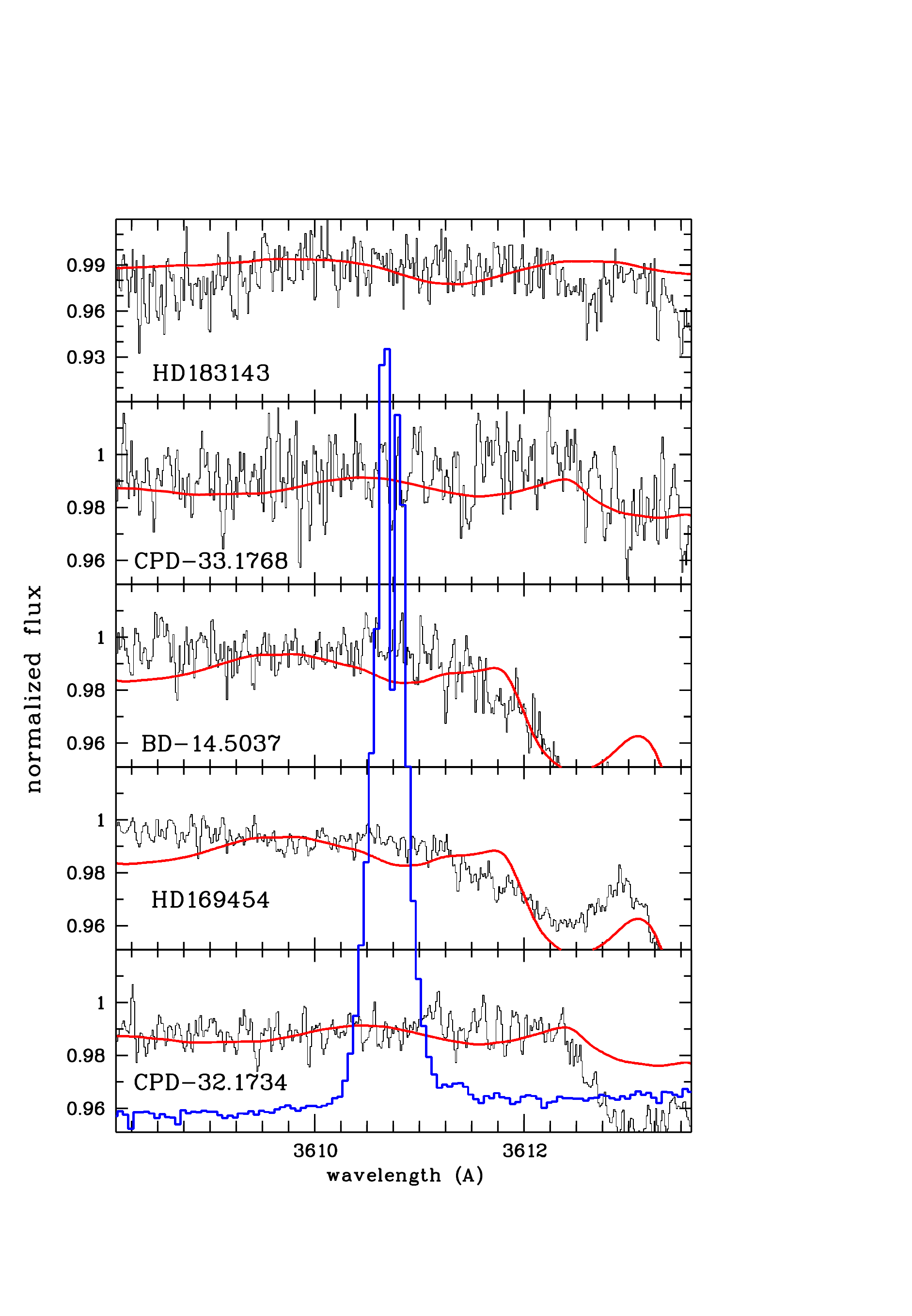}
      \caption{
Comparison between the $S_1 \leftarrow S_0$ origin band of anthracene measured in our laboratory
(plotted in blue, \citet{Staicu04} and the UVES spectra obtained toward our programme stars.
Red lines represent synthetic stellar spectra obtained from the models of
\citet{gummersbach}. All stellar spectra have been rebinned to a wavelength frame where
the detected interstellar absorption lines occur at an average radial velocity
of 0 km s$^{-1}$, compared with the respective rest wavelengths in air.
              }
         \label{anthracene}
   \end{figure}

   \begin{figure}
   \centering
   \includegraphics[angle=0,width=10cm]{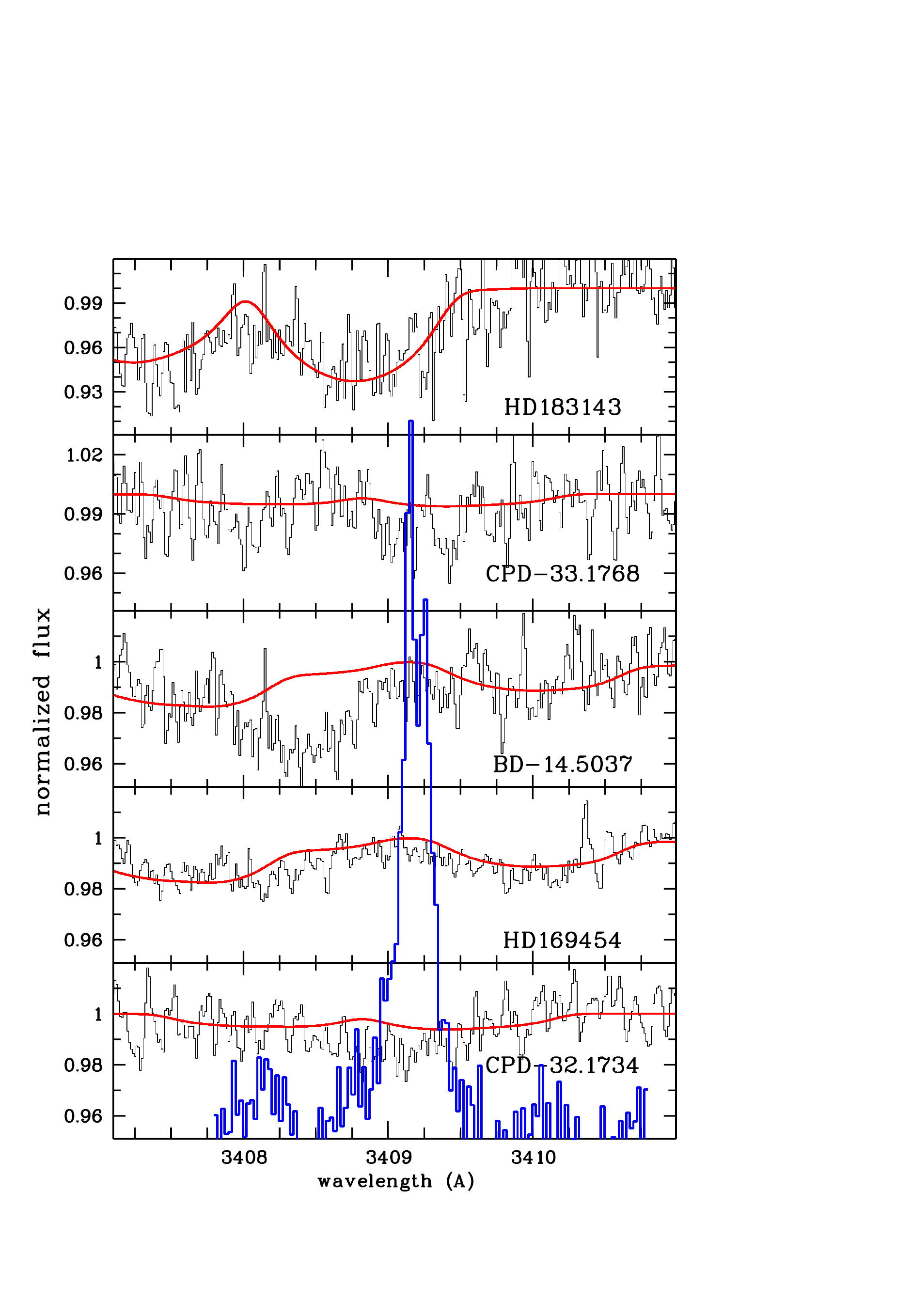}
      \caption{
Comparison of the weak $S_1 \leftarrow
S_0$ transition of phenanthrene (in blue) with UVES spectra of our
programme stars.  See  Fig.~\ref{anthracene}
for further details.
              }
         \label{phenanthrene}
   \end{figure}


\section{Observations and data reduction}
\label{observations}

We have obtained high-resolution spectra toward the heavily reddened supergiants
HD 169454 (B1.5Ia, $E_\mathrm{B-V}$ = 1.13 mag),
BD $-14^\circ 5037$ (B1.5Ia, $E_\mathrm{B-V}$ = 1.56 mag),
CPD $-32^\circ 1734$ (B1.5Ia, $E_\mathrm{B-V}$ = 1.22 mag),
CPD $-33^\circ 1768$ (B2.5I, $E_\mathrm{B-V}$ = 1.46 mag), and
HD 183143 (B7Iae, $E_\mathrm{B-V}$ = 1.27 mag),
using the UVES spectrograph at the VLT under programme 079.C-0597(A).
Total integration times were $14 \times 300$ sec for CPD $-32^\circ$1734, $11 \times 500$ sec for CPD $-33^\circ$1768,
$16 \times 250$ sec for BD $-14^\circ$5037, $20 \times 50$ sec for HD 169454,
and $6 \times 50$ sec for HD 183143. The spectra are reduced within the ESO MIDAS package following
standard procedures and are rebinned to a heliocentric wavelength scale.

We also use the archival UVES spectra obtained under the UVES Paranal Observatory Project
'A Library of High-Resolution Spectra of Stars across the Hertzsprung-Russell Diagram' \citep{bagnulo03}
produced under ESO Director Discretionary Time (DDT) program 266.D-5655(A).
{The database of \citet{bagnulo03} contains a large number of field stars with spectral types ranging from 
O to M and luminosity classes ranging from main sequence stars to supergiants. Here we select 
a few of the more reddened luminous O- and B-type stars with reddenings up to $E_\mathrm{B-V}$ = 1 mag,}
HD 97253 (O5.5III, $E_\mathrm{B-V} = 0.50$ mag),
HD 148937 (O6.5, $E_\mathrm{B-V} = 0.67$ mag),
HD 96917 (O8.5Ib, $E_\mathrm{B-V} = 0.37$ mag),
HD 76341 (O9Ib, $E_\mathrm{B-V} = 0.46$ mag),
HD 152003 (O9.7Iab, $E_\mathrm{B-V} = 0.64$ mag), and toward the reddened B-type supergiants
HD 112272 (B0.5Ia, $E_\mathrm{B-V} = 0.99$ mag),
HD 115363 (B1Ia, $E_\mathrm{B-V} = 0.82$ mag),
HD 152235 (B1Ia, $E_\mathrm{B-V} = 0.73$ mag), and
HD 148379 (B1.5Ia, $E_\mathrm{B-V} = 0.71$ mag),
{Values of $E_\mathrm{B-V}$ are from \citet{kaz2010}.} 
The archival spectrum of HD 169454 is used for consistency checks.

{The archive contains several stars with low reddening, which we use to identify stellar
absorption lines. We select the spectra toward HD 29138 (B1 Iab), HD 58978 (B1 II), and Rigel (B8 Iab), 
which are characterized by very faint or absent interstellar molecular absorption lines of
CH 4300 $\AA$, CH$^+$ 4232 $\AA$, and CN 3875 $\AA$, and faint or absent
\ion{Ti}{II} 3229 $\AA$ and \ion{Na}{I} 3302 $\AA$ absorption lines.}
We use the models of \citet{gummersbach} to assess the stellar continua of our programme stars.
These models are based on the LTE static plane-parallel line-blanketed ATLAS9 model
{\it atmosphere} of \citet{kurucz91} and were developed for main-sequence stars with spectral
types ranging from O9V to B5V. The stars in our sample are supergiants in general,
some of them with strong stellar winds. Stellar parameters such as gravity, rotation,
and metallicity which affect the width and the depth of the photospheric absorption lines
may thus deviate significantly from those adopted in the models used here.
Because of this, the models of \citet{gummersbach} may not reproduce the stellar spectra
presented here in all details, yet the models are still extremely useful to distinguish
among stellar and potential interstellar absorption features in our UVES spectra.

\section{Results}
\label{results}

\subsection{Comparison of astrophysical and laboratory spectra}
\label{comparisons}

\begin{figure}
   \centering
   \includegraphics[angle=0,width=10cm]{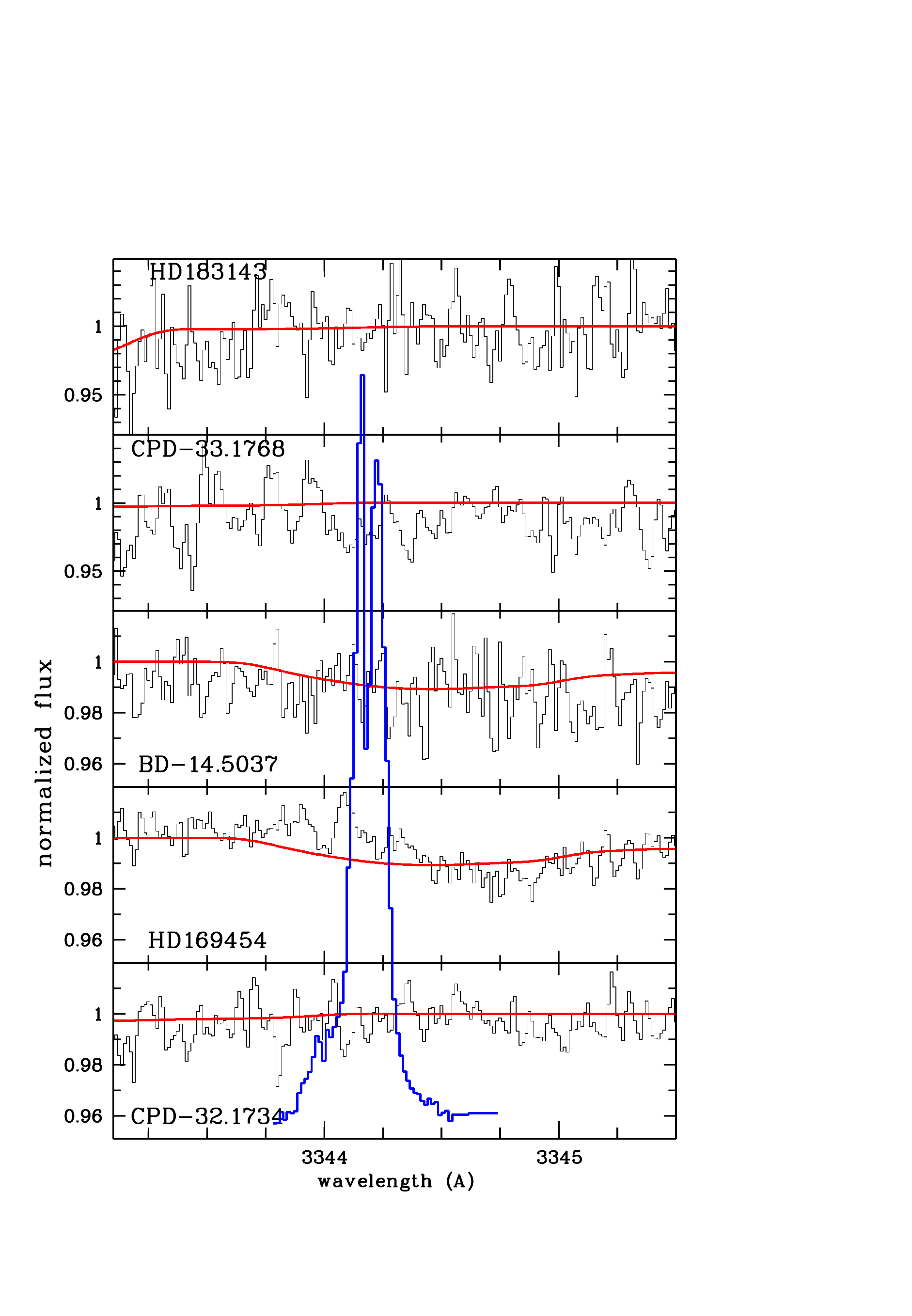}
      \caption{
Comparison of the origin band of the $S_1 \leftarrow S_0$ transition of
{2,3-benzofluorene}
with the UVES spectra of our programme stars.
See Fig.~\ref{anthracene} for further details.
              }
         \label{benzofluorene1}
   \end{figure}

\begin{figure}
   \centering
   \includegraphics[angle=0,width=10cm]{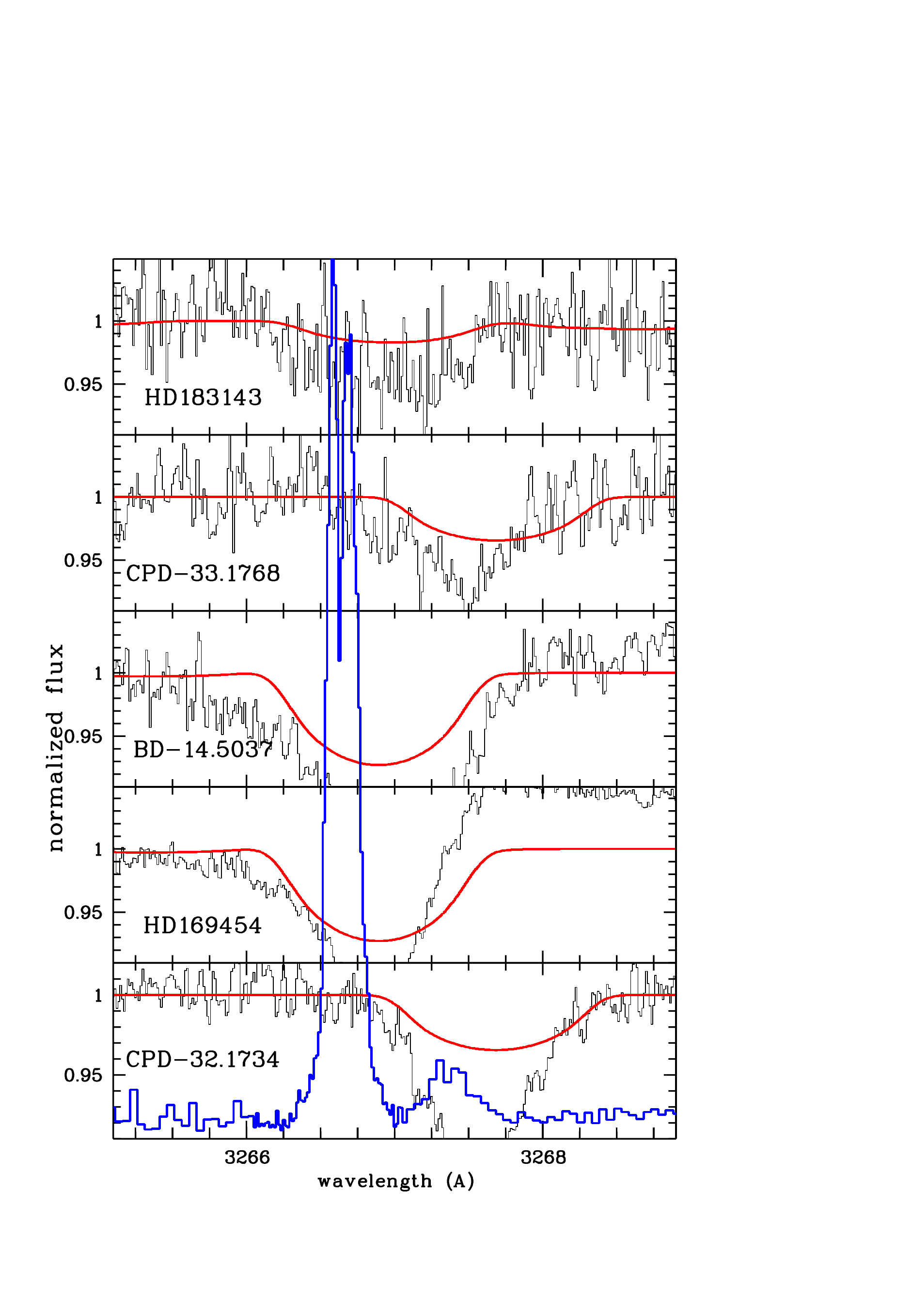}
      \caption{
Comparison of laboratory data of 2,3-benzofluorene (in blue) with
UVES spectra of our programme stars. The blue curve represents an electronic transition involving the
additional excitation of a vibration with an energy of 700 cm$^{-1}$ \citep{Staicu08}.
See Fig.~\ref{anthracene} for further details.
              }
         \label{benzofluorene2}
   \end{figure}

\begin{figure}
   \centering
   \includegraphics[angle=0,width=10cm]{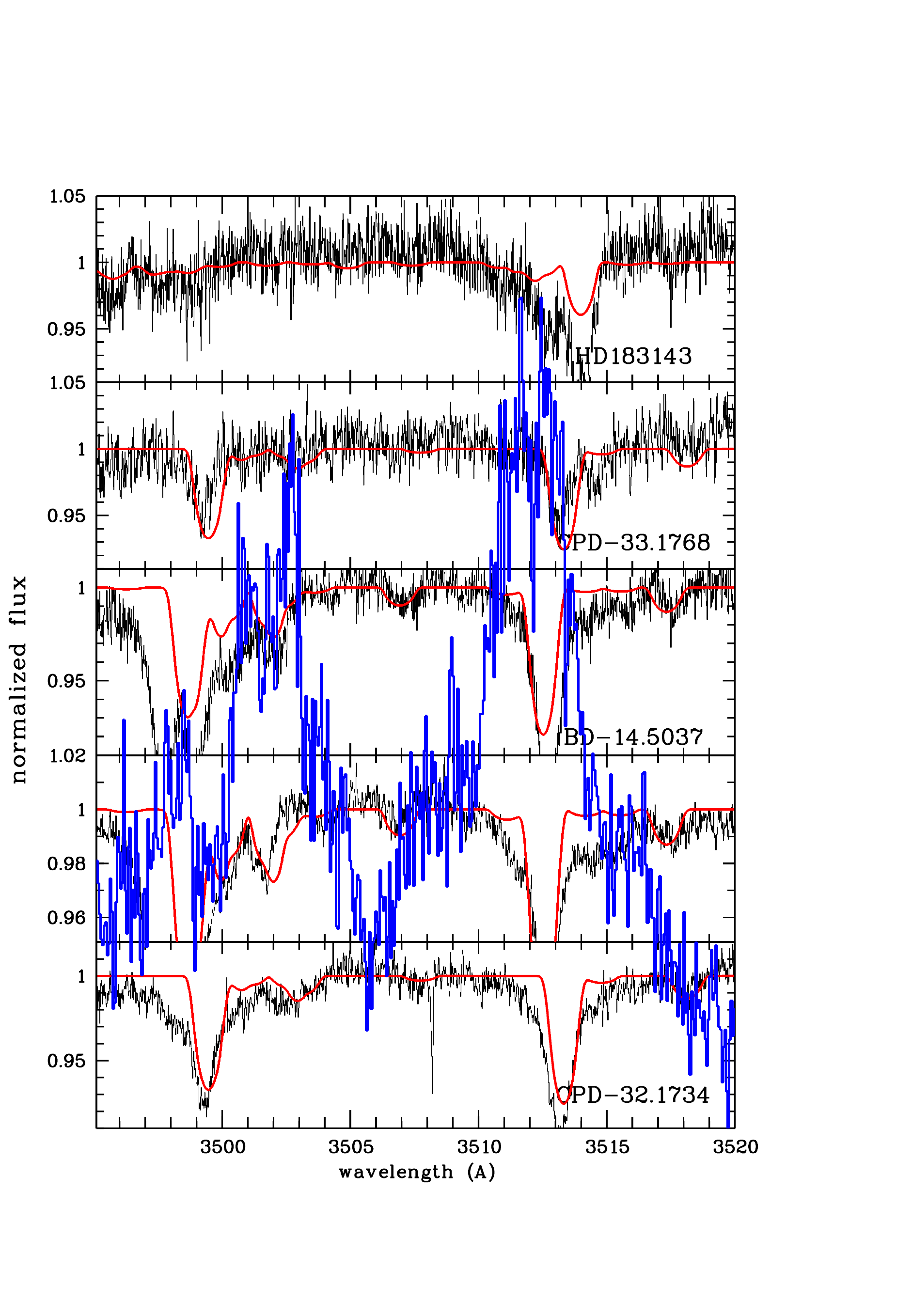}
      \caption{
Comparison of laboratory data of benzo[ghi]perylene (in blue) with
UVES spectra of our programme stars. The laboratory spectrum shows two vibrational bands in
the $S_2\leftarrow S_0$ transition \citep{Tan05}.
See Fig.~\ref{anthracene} for further details.
              }
         \label{benzoperylene}
   \end{figure}

\begin{figure}
   \centering
   \includegraphics[angle=0,width=10cm]{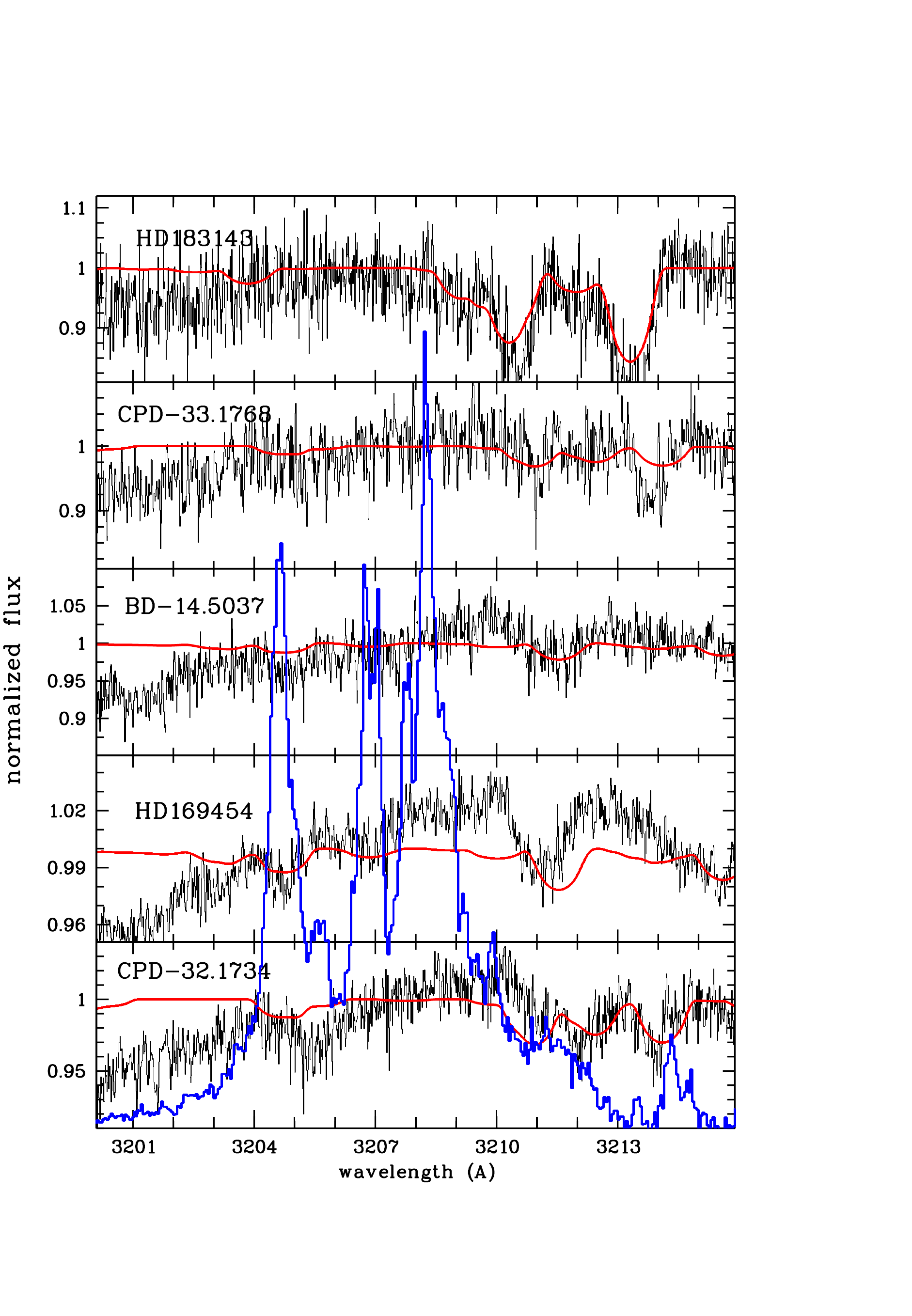}
      \caption{
UVES spectra in the region where the origin of the $S_2 \leftarrow S_0$ transition of pyrene
occurs \citep{Rouille04}. See Fig.~\ref{anthracene} for further details.
              }
         \label{pyrene1}
   \end{figure}

\begin{figure}
   \centering
   \includegraphics[angle=0,width=10cm]{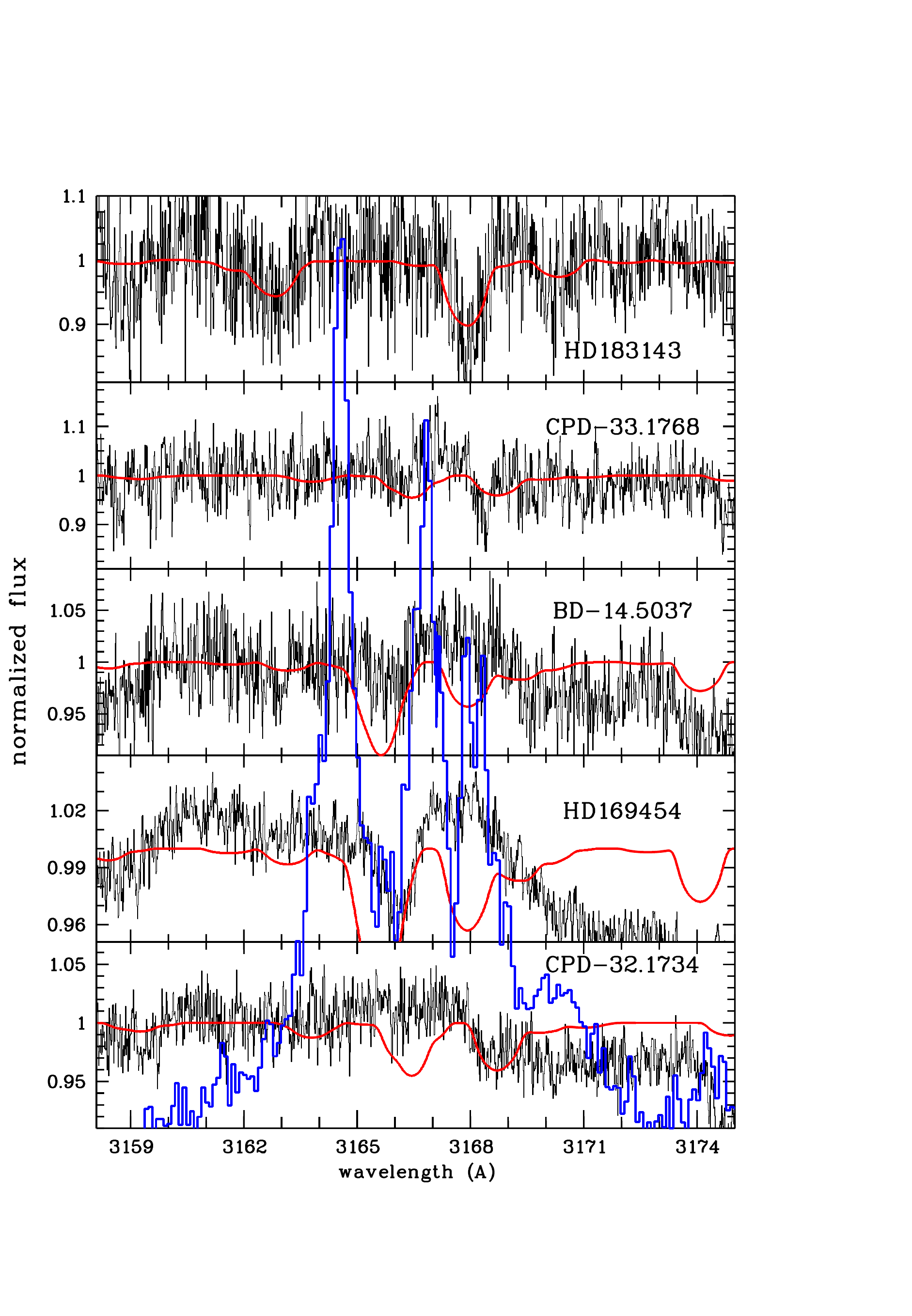}
      \caption{
UVES spectra of another region of the $S_2 \leftarrow S_0$ transition of pyrene
involving electronic and vibrational excitation \citep{Rouille04}.
See Fig.~\ref{anthracene} for further details.
              }
         \label{pyrene2}
   \end{figure}

\begin{figure*}
   \centering
   \includegraphics[angle=0,width=10cm]{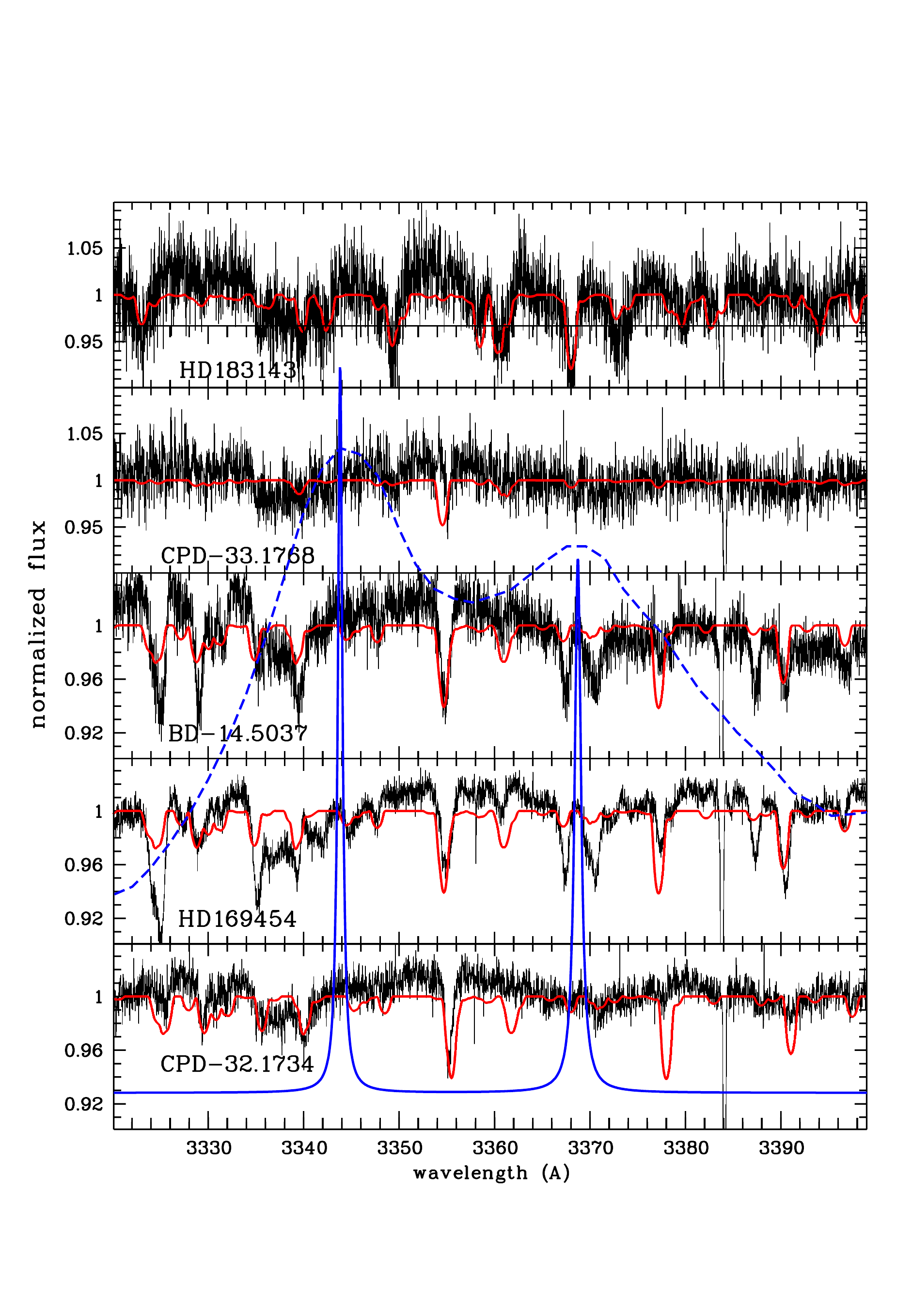}
      \caption{
UVES spectra covering the region where the strongest absorption peaks in the $S_\beta
\leftarrow S_0$ transition of hexabenzocoronene are expected. The dashed blue curve
represents the hexabenzocoronene spectrum as measured in a Ne matrix and shifted to
the expected gas phase position following the method described in Sect. 2.2. The two
narrow bands, plotted as solid blue curve, are synthetic low-temperature gas phase
profiles drawn at their predicted gas phase positions (see text).
The sharp absorption feature at 3384 $\AA$ is due to interstellar \ion{Ti}{II}.
See Fig.~\ref{anthracene} for further details.
              }
         \label{hexa1}
   \end{figure*}

In the standard reduction scheme adopted here (cf. Sect.~\ref{observations}), the
spectra obtained with UVES are rebinned to a heliocentric
wavelength scale. In order to compare the astronomical spectra with our laboratory spectra,
we proceed as follows. The astronomical spectra contain a number of
interstellar absorption lines from diatomic molecules (cf. Sect.~\ref{islines}
and Table~\ref{ismolecules}) which we use to infer average heliocentric velocities
$V_\mathrm{ave}$ of
$-9.8$ km s$^{-1}$,
$-7.9$ km s$^{-1}$,
$+36.8$ km s$^{-1}$, and
$+35.9$ km s$^{-1}$ of the molecular material
toward HD 169454, BD $-14^\circ 5037$, CPD $-33^\circ 1768$, and CPD $-32^\circ1734$, respectively.
We now shift the stellar spectra by the respective values of $-V_\mathrm{ave}$ and plot
the modified spectra, together with the laboratory absorption profiles of the PAHs studied here,
in Figs.~\ref{anthracene} - \ref{hexa1}.  Assuming that the sought PAH absorbers reside in the
molecular material traced by the diatomic molecules detected here (cf. Table~\ref{ismolecules}),
corresponding interstellar absorption features must agree in
position and in spectral profile with the laboratory profiles plotted in Figs.~\ref{anthracene}
- \ref{hexa1}. Note that a velocity dispersion of up to 4 km s$^{-1}$ among the various
interstellar absorbers toward a given line of sight would translate into a wavelength spread of
about one resolution element only, which is negligible.

Figure~\ref{anthracene} presents the comparison of the origin band
of the $S_1 \leftarrow S_0$ transition of anthracene, which is
plotted in blue \citep{Staicu04}, with the UVES spectra of our five programme stars.
Red lines represent synthetic stellar profiles obtained from the models of
\citet{gummersbach}. Toward HD 169454 and BD $-14^\circ5037$,
the expected absorption of anthracene agrees in wavelength with a faint
photospheric \ion{Fe}{III} (3611.7 $\AA$) line. Stronger stellar absorption
lines at 3612.4 $\AA$ and 3613.6 $\AA$  correspond to \ion{Al}{III} and \ion{He}{I}, respectively.
Toward the two B1.5Ia supergiants HD 169454 and BD $-14^\circ5037$, the stellar
continuum near the expected absorption of anthracene is not
very well reproduced by the models of \citet{gummersbach}, yet the discrepancies are
assigned to effects of stellar parameters such as gravity and metallicity.
It can thus be concluded that the spectra shown in
Fig.~\ref{anthracene} do not reveal signatures of interstellar anthracene absorptions
at 3610.74 $\AA$. The observations may nevertheless be used to infer upper limits in
the column density and in the fractional abundance of anthracene
(cf.  Sect.~\ref{sectionupperlimits}) toward our lines of sight.

The laboratory spectrum of the $S_1 \leftarrow S_0$ origin band of
phenanthrene {at 3409.21 $\AA$ is given} in
Fig.~\ref{phenanthrene} together with the five UVES spectra. The
transition is significantly weaker than the $S_2 \leftarrow S_0$
origin band, yet the latter occurs  at 2826.42 $\AA$ which is not
accessible from the ground. The corresponding spectral region
toward the B7Ia supergiant HD 183143 shown in
Fig.~\ref{phenanthrene} is dominated by stellar \ion{Ni}{II}
(3407.3 $\AA$) and \ion{Cr}{II} (3408.8 $\AA$) absorption lines
which are well reproduced by the stellar models. The hotter
atmospheres of HD 169454 and BD $-14^\circ5037$ contain faint
absorptions arising from \ion{Ne}{II} (3406.9 $\AA$) and
\ion{O}{II} (3407.3 $\AA$, 3409.8 $\AA$). Toward BD
$-14^\circ5037$, an absorption feature occurs near 3408 $\AA$
which is not reproduced  by the stellar models. The stellar
continuum of the unreddened star HD 148379, which is a B1.5 Ia
supergiant as well, is featureless in the spectral region shown in
Fig.~\ref{phenanthrene}. It is thus tempting to assign the
absorption feature near $\lambda_\mathrm{air} = 3408.3\ \AA$
toward BD $-14^\circ$ 5037 to a new diffuse interstellar band, yet
its origin from phenanthrene is rejected. The absorption feature
is significantly broader than the phenanthrene band and it would
be blue-shifted by some 100 km s$^{-1}$ if due to phenanthrene.
BD$-14^\circ$5037 and HD 169454 are members of the Sct OB3
association where high-velocity gas has been detected
\citep{federman92}. The high-velocity gas, redshifted by some 100
km s$^{-1}$, has been detected toward HD 169454 and a few other
stars in the Sct OB3 association, but not toward
BD$\-14^\circ$5037. Its origin has been assigned to an expanding
filamentary shell, probably due to a SN explosion in Sct OB3.

Figures~\ref{benzofluorene1} and \ref{benzofluorene2} present the comparison of the laboratory
spectra of 2,3-benzofluorene with our  UVES spectra.
Figure~\ref{benzofluorene1} displays the vibrationless origin band of the $S_1
\leftarrow S_0$ transition of 2,3-benzofluorene at 3344.16 {\AA}.
The hotter stars such as HD 169454 show faint stellar absorption lines arising
from \ion{Ne}{II} (3344.4 $\AA$ and 3345.5 $\AA$) and \ion{Ar}{III} (3344.7 $\AA$) which
merge into a shallow and unresolved absorption feature. The relatively featureless stellar
continua of our programme stars in this wavelength region favor the search of the relatively
narrow $S_1 \leftarrow S_0$ transition of 2,3-benzofluorene, yet none of the spectra
show a corresponding absorption feature near 3344.16 $\AA$.  Figure~\ref{benzofluorene2} shows a
weaker band of 2,3-benzofluorene located at 3266.63 {\AA}. That band involves
an additional vibrational excitation. A strong
stellar \ion{Fe}{III} absorption (3266.9 $\AA$) is present in all spectra, and it
occurs at the expected position of the 2,3-benzofluorene absorption. The \ion{Fe}{III} line
toward HD 169454 exhibits a strong P Cygni profile. The adopted models do not reproduce
the depth nor the profiles of the \ion{Fe}{III} line
toward CPD $-32^\circ1734$, HD 169454, and BD $-14^\circ 5037$. It can thus not be ruled
out that additional absorption due to 2,3-benzofluorene occurs near 3266.63 $\AA$,
yet this is unlikely because of the absence of the stronger band at 3344.16 $\AA$.

Part of the laboratory spectrum of benzo[ghi]perylene showing two broad vibrational features at
3501.76 $\AA$ and 3512.15 $\AA$ \citep{Tan05}
and the corresponding UVES spectra are shown in Fig.~\ref{benzoperylene}.
The two vibrational bands belong to the $S_2 \leftarrow S_0$ transition.
The coincidence of various stellar absorption lines in this wavelength region
renders the detection of the relatively broad absorption bands of benzo[ghi]perylene
very difficult. The spectra of the hotter stars show
\ion{He}{I} absorptions near 3498.7 $\AA$, 3502.4 $\AA$, and 3512.5 $\AA$, which coincide with
the absorption wavelengths of benzo[ghi]perylene. Other photospheric lines in this region
are \ion{Ne}{II} (3503.6 $\AA$) and \ion{Cr}{VII} (3498.9 $\AA$). The cooler stars show
\ion{Ni}{II} (3514.0 $\AA$) absorption with shallow \ion{Cr}{II} (3511.8 $\AA$) and
\ion{He}{I} (3512.5 $\AA$).  The stellar models do not reproduce the observed spectra near
these wavelengths with sufficient accuracy,
and some residuals exist near 3514 $\AA$ where benzo[ghi]perylene absorptions are
expected. The unreddened archival stars of similar spectral type show identical
photospheric line profiles with broad wings, which leads us to reject possible contributions
from benzo[ghi]perylene to the residuals seen near 3514 $\AA$. The
strong benzo[ghi]perylene origin band of the $S_2 \leftarrow
S_0$ transition occurs near 3685 $\AA$ and falls into a spectral region which is dominated by strong
photospheric hydrogen absorption lines converging toward the Balmer $\infty \rightarrow 2$
series limit at 3646 $\AA$. Because of the strong photospheric Balmer lines, a search of
the origin band of the $S_2 \leftarrow S_0$ transition of benzo[ghi]perylene toward early
type supergiants is hopeless.

Our UVES spectra are rather noisy indeed at wavelengths below 3210 {\AA} where the $S_2 \leftarrow S_0$ absorptions
of pyrene occur. The signal-to-noise ratio of our
spectra is smaller than $100$ in general, yet the spectra are still useful as
the pyrene transitions exhibit large oscillator strengths.
As mentioned in Sect.~\ref{method}, the interaction of the $S_2$ state of pyrene
with the vibrational manifold of the $S_1$ state gives rise to groups of bands instead of isolated features.
Figure~\ref{pyrene1} shows
the group of bands of strong intensities that appear between 3204 -- 3209 {\AA} in the place of the origin
band of $S_2 \leftarrow S_0$ \citep{Rouille04}. The UVES data reveal faint stellar absorption
lines due to \ion{Fe}{III} (3204.8 $\AA$) and \ion{He}{I} (3211.6 $\AA$) in the spectra of the hotter stars.
The stellar spectra reveal no signature that correspond to the group of pyrene bands plotted in
Fig.~\ref{pyrene1}.
Figure~\ref{pyrene2} compares another group of bands
between 3163 -- 3169 $\AA$ which correspond to the excitation of an in-plane CCC bending mode together
with the $S_2$ electronic state of pyrene. These bands coincide with the stellar \ion{C}{II} (3165.5, 3166.0, and
3167.9 $\AA$) and \ion{Si}{IV} (3165.7 $\AA$) absorption lines, which makes a detection of pyrene in this
wavelength region difficult, even in spectra of significantly higher S/N. Thus, the 3204 -- 3209 $\AA$
region shown in Fig.~\ref{pyrene1} with its few and faint stellar lines appears more promising
to search for interstellar pyrene. In addition, the pyrene bands in this wavelength region are stronger than
between 3163 -- 3169 $\AA$.

Figure~\ref{hexa1} presents the spectral region where the strongest bands of the $S_\beta
\leftarrow S_0$ transition of hexabenzocoronene is expected. The spectral profile
for hexabenzocoronene (dashed blue line) was measured in a Ne
matrix and shifted to the expected gas phase position following the method
described in Sect.~\ref{lab-wavelengths}. We are not able to provide absorption cross sections
for hexabenzocoronene because our laboratory data does not reveal the number of individual
gas phase bands that form the broad structures seen in Fig.~\ref{hexa1}.
A simulated low-temperature gas phase spectrum (solid blue line)
that can be compared with the UVES spectra is also shown in
Fig.~\ref{hexa1}. It represents two bands at the gas phase
positions predicted for the strongest peaks in the $S_\beta
\leftarrow S_0$ transition of hexabenzocoronene, i.e. at 3344 and
3369 {\AA}. The synthetic bands are represented in terms of
Lorentzian profiles, where the ratios of their widths and their
heights are obtained from the spectrum of hexabenzocoronene
measured in the Ne matrix. We have chosen to fix the widths to 5
and 7 cm$^{-1}$ because these values are typical of the bandwidths
observed in the jet-cooled spectrum of the $S_2 \leftarrow S_0$
transition of benzo[ghi]perylene \citep{Tan05}. This transition is
affected by an interaction between electronic states \citep{Tan05}
like the $S_\beta \leftarrow S_0$ transition of hexabenzocoronene
\citep{Rouille2009}. As stated in Sec.~\ref{lab-wavelengths},
groups of bands, not just a pair, are actually expected in the
low-temperature gas phase spectrum of hexabenzocoronene in the
region covered by Fig.~\ref{hexa1}. Our UVES spectra do not reveal
features resembling the simulated bands of hexabenzocoronene.

\begin{figure}
   \centering
   \includegraphics[angle=0,width=10cm]{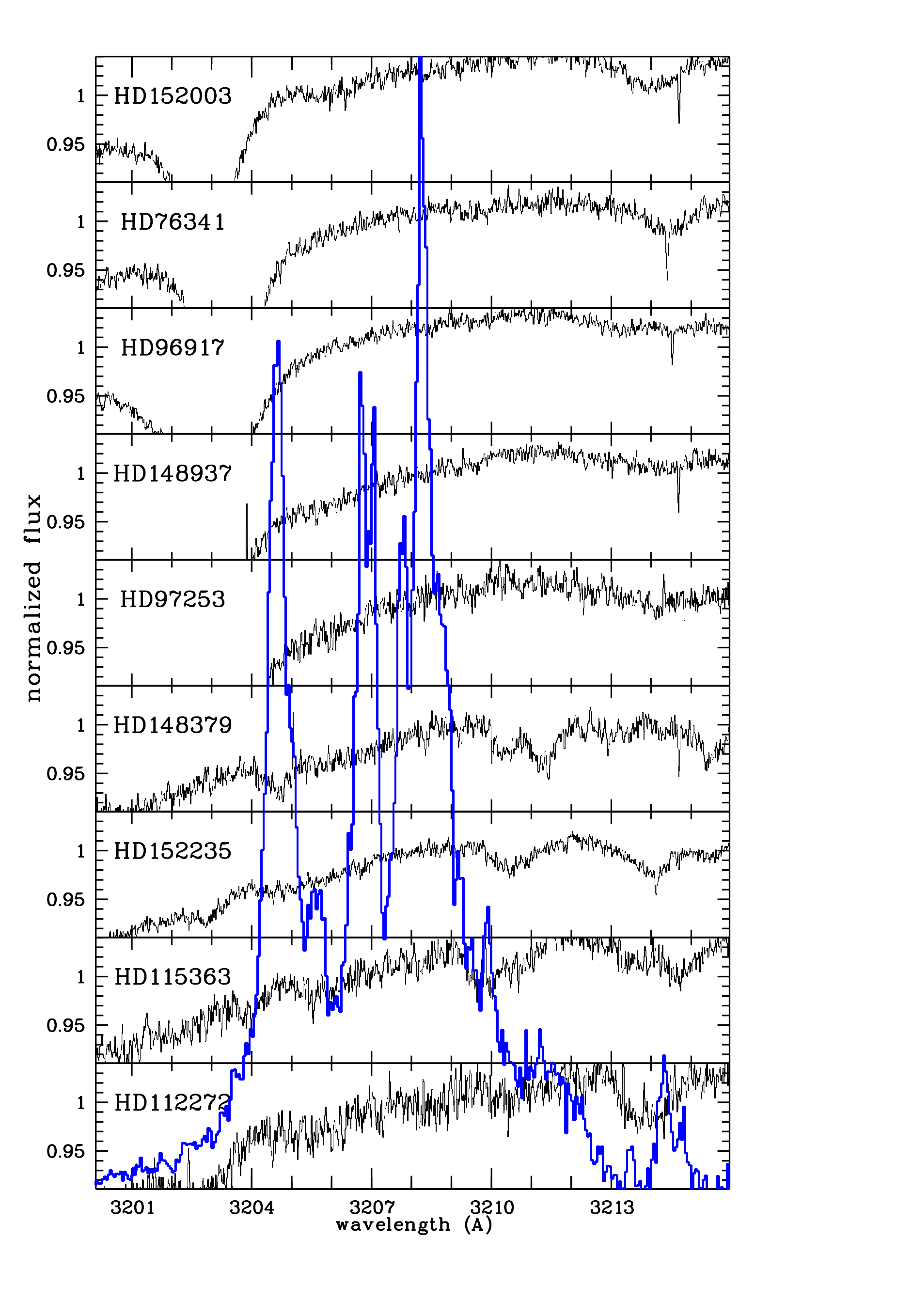}
      \caption{
Archival spectra covering the region where the origin of the $S_2 \leftarrow S_0$ transition of pyrene
occurs \citep{Rouille04}. See Fig.~\ref{anthracene} for further details.
              }
         \label{pyrene1arx}
   \end{figure}

\begin{figure}
   \centering
   \includegraphics[angle=0,width=10cm]{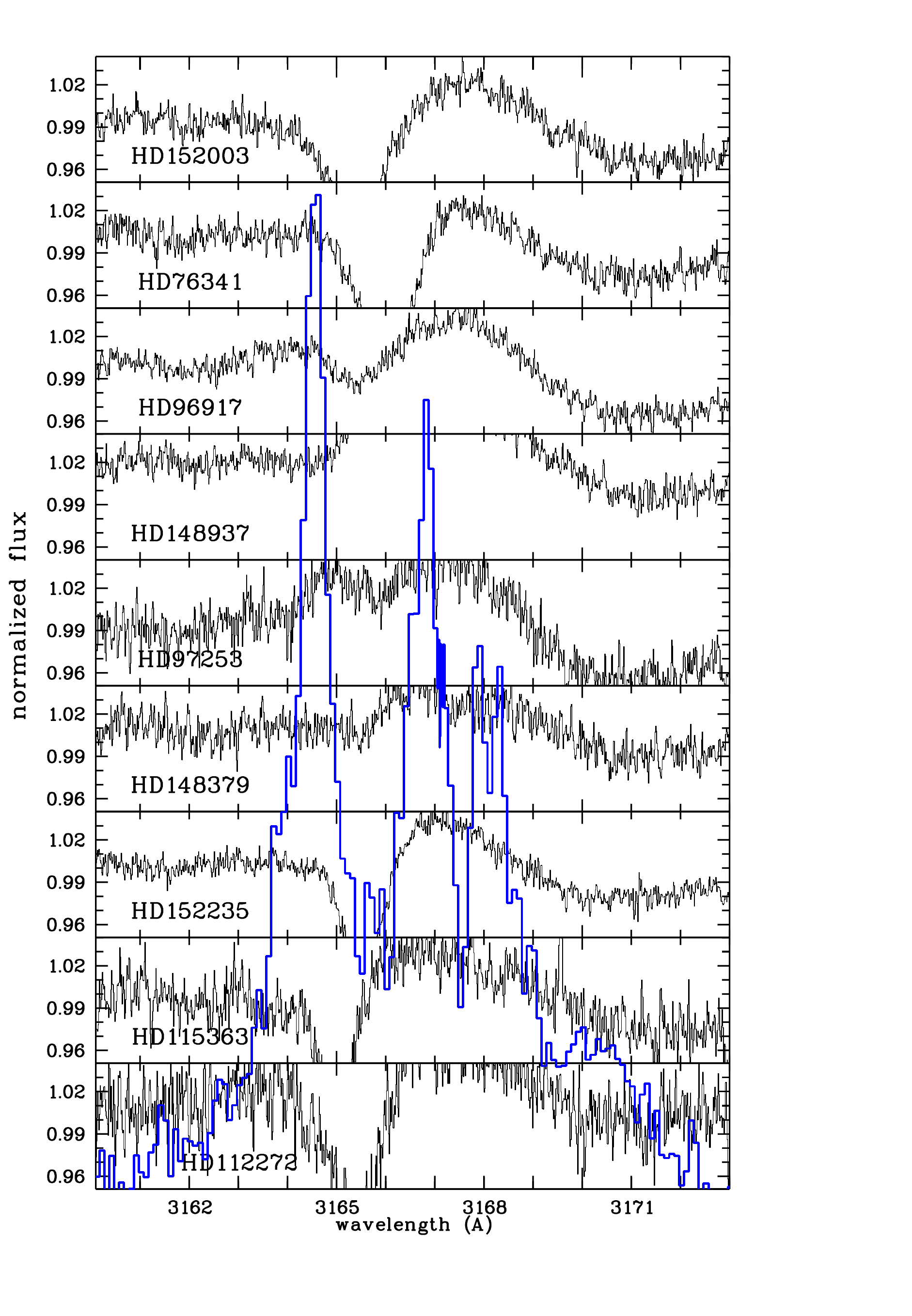}
      \caption{
Archival spectra covering a $S_2 \leftarrow S_0$ transition of pyrene
involving electronic and vibrational excitation \citep{Rouille04}.
See Fig.~\ref{anthracene} for further details.
              }
         \label{pyrene2arx}
   \end{figure}

We have investigated all archival spectra for potential PAH absorptions. The archival spectra
were obtained toward stars with significantly lower reddening, in general, and consequently,
the S/N ratio in the blue part of the spectrum is higher than in our spectra.
None of the archival spectra reveal traces of the PAHs studied here. This is illustrated
in Figs.~\ref{pyrene1arx} and \ref{pyrene2arx}, which cover the spectral region where the
two groups of bands of pyrene studied here occur. These two
pyrene bands have relatively large oscillator strengths (cf. Sect.~\ref{method} and
Table~\ref{tab:cross-section}), so relatively small column densities of pyrene become detectable.
None of the archival spectra show signatures of interstellar pyrene, however. The upper limits
in the equivalent widths of pyrene are nevertheless used to constrain the abundance of pyrene
in translucent material, as will be demonstrated in the following section.

\subsection{Constraints of PAH column densities in translucent material}
\label{sectionupperlimits}

Our UVES spectra together with the laboratory data may be used to infer upper limits in the column densities of
anthracene, phenanthrene, pyrene, 2,3-benzofluorene, and benzo[ghi]perylene. In cases where
a sought absorption band is not detected,
an upper limit in the equivalent width $W_\lambda$ of the absorption band is given by
$W_{\lambda} =  3 \sqrt{M} \Delta\lambda / \mathrm{(S/N)}$,
where $\Delta \lambda$ is the spectral width of a resolution element, $M$ is the number
of resolution elements over which the sought absorption band spreads, and
S/N is the signal-to-noise ratio near the expected absorption feature (see \citet{bohlin}
for a comprehensive treatment of error propagation in absorption line measurements).
The equivalent width $\mathrm W_\lambda$ of an interstellar absorption line
is defined in terms of
      {$W_\lambda = \int{(F_\mathrm{c} - F_\lambda) / F_\mathrm{c}\
      \mathrm{d}\lambda}$}
where $F_\lambda$ is the measured flux level at wavelength $\lambda$ and where
$F_\mathrm{c}$ is the flux in the continuum adjacent to the absorption line.
The equivalent width is related to the
optical depth $\tau(\lambda)$ according to
\begin{equation}
      W_{\lambda} = \int{(1-e^{-\tau(\lambda)}) \mathrm{d}\lambda},
\label{eqno1}
\end{equation}
which, for low optical depths $\tau(\lambda)<<1$, is expressed as
\begin{equation}
      W_{\lambda} = \int\tau(\lambda)\mathrm{d}\lambda = N \int{\sigma(\lambda)} \mathrm{d}\lambda.
\label{eqno2}
\end{equation}
$N$ is the number of absorbers per cm$^{2}$ and $\sigma(\lambda)$
is the absorption cross section.

Our UVES spectra were obtained with a spectral resolution of
$R = \lambda / \Delta \lambda = 80\,000$. For simplicity, we
adopt a constant width of $\Delta \lambda = 45\ \mathrm{m} \AA$ per resolution element
over the entire wavelength range studied here.
Table~\ref{upperlimits} summarizes the central wavelengths and
the corresponding values of $3 \sqrt{ M} \Delta \lambda$ for each of the sought absorption bands,
and the corresponding signal-to-noise ratio S/N, the upper limits in $W_\lambda$, and the upper
limits in the column densities $N$ toward
HD 169454, BD $-14^\circ5037$, CPD $-33^\circ1768$, CPD $-32^\circ1734$, and HD 183143.

\begin{table*}
\caption{Upper limits of large molecules toward the lines of sight studied here.}
\label{upperlimits}
\centering
\begin{tabular}{lrlllllllllllllllllllll}
\hline\hline
& & \multicolumn{3}{l}{HD 169454} &
\multicolumn{3}{l}{BD $-14^\circ5037$} &
\multicolumn{3}{l}{CPD $-33^\circ1768$} &
\multicolumn{3}{l}{CPD $-32^\circ1734$} &
\multicolumn{3}{l}{HD 183143} \\
$\lambda_\mathrm{air}$ & $3 \sqrt{M} \Delta \lambda $
& S/N & $W_\lambda$ & $N_\mathrm{max}^a$
& S/N & $W_\lambda$ & $N_\mathrm{max}^a$
& S/N & $W_\lambda$ & $N_\mathrm{max}^a$
& S/N & $W_\lambda$ & $N_\mathrm{max}^a$
& S/N & $W_\lambda$ & $N_\mathrm{max}^a$ \\
\hline
\multicolumn{2}{l}{anthracene}         \\
3610.74 & 371 & 230 & 1.6 & 0.8& 113 & 3.3 & 1.7& 81 & 4.6 & 2.4 & 143 & 2.6 & 1.4& 69 & 5.4 & 2.8\\
\multicolumn{2}{l}{phenanthrene}       \\
3409.21 & 284 & 165 & 1.7 & 105&  97 & 2.9 & 180& 62 & 4.6 & 285 & 145 & 1.9 & 118& 53 & 5.3 & 330\\
\multicolumn{2}{l}{pyrene}             \\
3208.22 & 1493& 70  & 21  & 2.3&  37 & 40  & 4.4& 23 & 65  & 7.2 & 51  & 29  & 3.2& 23 & 65  & 7.1\\
\multicolumn{2}{l}{2,3-benzofluorene}  \\
3344.16 & 270 & 117 & 2.3 & 2.7&  88 & 3.1 & 3.7& 47 & 5.7 &6.8 & 117 & 2.3 & 2.8& 41 & 6.6 & 7.9 \\
\multicolumn{2}{l}{benzo[ghi]perylene} \\
3512.15 & 1822& 206 & 8.8 & 8.8 & 125 & 14.6 & 14.6 & 82 & 22 & 22 & 165 & 11 & 11 & 76 & 24 & 24 \\
3501.76 & 1465 & 206 & 7.1 & 12.8 & 125 & 11.7 & 21 & 82 & 18 & 32 & 165 & 8.9 & 16 & 76 & 19 & 34\\
\hline
\multicolumn{6}{l}{$^a$ Upper limits $N_\mathrm{max}$ in units of $10^{12}$\ cm$^{-2}$} \\
\end{tabular}
\end{table*}

Our data indicates maximal column densities of anthracene, pyrene, and benzo[ghi]perylene
of a few $10^{12}$\ cm$^{-2}$ toward the heavily reddened programme stars of our UVES sample.
Toward the stars retrieved from the archive, upper column densities in pyrene are below values
of $10^{12}$ cm$^{-2}$.  Using the relations of \citet{mattila86}, the upper limits in the PAH
column densities and the observed CH column densities (cf. Sect.~\ref{islines})
 are used to  infer upper limits in the fractional abundances of
$f(\mathrm{PAH}) = N(\mathrm{PAH})/N(\mathrm{H}) = 1.1 \times
10^{-8} N(\mathrm{PAH})/N(\mathrm{CH})$. Our observations indicate
values of $f$(anthracene) $\le 1.5 - 2 \times 10^{-10}$ toward
HD169454, BD$-14^\circ5037$, CPD$-32^\circ1734$, and
CPD$-33^\circ1768$, and $f$(anthracene) $\le 10^{-9}$ toward HD
183143. The upper limits in the fractional abundances of pyrene
and 2,3-benzofluorene are $\le 5 \times 10^{-10}$ toward HD 169454
and BD$-14^\circ5037$ and $\le 3.5 \times 10^{-9}$ toward the
remaining three stars. Fractional abundances of pyrene toward the
more diffuse lines of sight traced by the archival stars are below
a few times $10^{-10}$. For phenanthrene, which has a relatively
small oscillator strength in the transitions studied here (cf.
Table~\ref{tab:cross-section}), upper limits in the fractional
abundances are a few $10^{-8}$. {Our values may
be compared to fractional abundances relative to H$_2$ derived for
other complex molecules from radio telescope observations.
\citet{Marcelino07} report a relative abundance of $4 \times
10^{-9}$ for propylene detected toward the dark cloud TMC-1 while
\citet{Belloche09} derived abundances for ethyl formate and
$n-$propyl cyanide observed toward Sgr B2(N)  of $3.6 \times
10^{-9}$ and $1.0 \times 10^{-9}$, respectively.}

\subsection{Interstellar molecular absorption lines}
\label{islines}
\label{molec}
Our UVES spectra contain various interstellar
absorption lines which are  presented for consistency checks with previous results.

\begin{figure}
   \centering
   \includegraphics[angle=0,width=10cm]{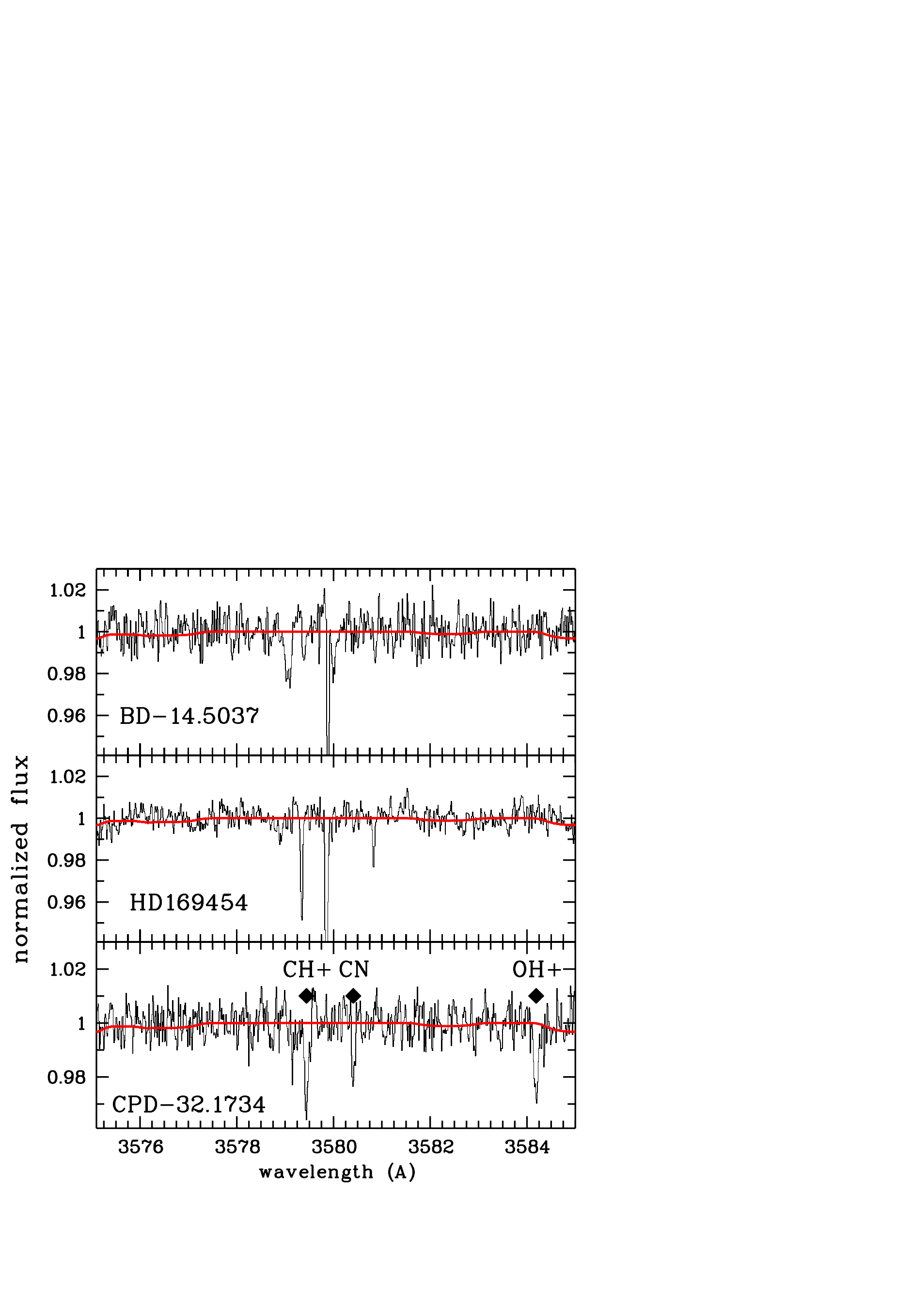}
      \caption{
Detection of the interstellar OH$^+$ radical in the spectrum toward CPD $-32^\circ$1734. Spectra
toward HD 169454 and BD $-14^\circ$5037
are shown as well. All spectra are normalized to a continuum level of 1.0 and rebinned to
a heliocentric wavelength scale. The OH$^+$, CH$^+$, and CN absorption lines toward
CPD $-32^\circ$1734 are explicitly identified. Synthetic stellar spectra obtained from
the models of \citet{gummersbach} are shown in red.
              }
         \label{ohp}
   \end{figure}

{\bf OH$^+$.}
Our spectrum obtained toward CPD $-32^\circ1734$ contains an
absorption line near 3584 $\AA$ which we assign to the interstellar
$R_{11}(0)\ (\lambda_\mathrm{air} = 3583.769\ \AA$)
line of the (0,0) band of the $\mathrm {A^3\Pi_i - X^3\Sigma^-}$ system of
OH$^+$. The relevant part of the spectrum is
reproduced in Fig.~\ref{ohp}, together with the spectra of
HD 169454 and BD $-14^\circ5037$.
Using the band oscillator strength of $f_{00} = 3.0764 \times 10^{-3}$ given by
\citet{almeida81},
we convert the measured equivalent width of $W_\lambda = 4\ \mathrm{m}\AA$ into an
OH$^+$ column density of $N$(OH$^+$) = $11 \times 10^{12}$\ cm$^{-2}$.
The OH$^+$ absorption line appears at a  heliocentric velocity of + 35.4 km s$^{-1}$
which is consistent with the average velocities of interstellar CN ($V_\mathrm{hel} = +37.5$ km s$^{-1}$),
CH$^+$ ($V_\mathrm{hel} = +35.6$ km s$^{-1}$), and CH ($V_\mathrm{hel} = +35.6$ km s$^{-1}$)
toward that line of sight (cf. Table~\ref{ismolecules}).

The OH$^+$ radical has only recently been discovered in the interstellar medium.
\citet{wyrowski10} detected the OH$^+$ $N=1-0$, $J=0-1$ absorption line toward the strong continuum
source Sgr B2 using APEX at sub-mm wavelengths, and \citet{gerin10} report the detection
of OH$^+$ absorption toward G10.6-04 (W31C) using HIFI on HERSCHEL.
Optical absorption lines arising from interstellar OH$^+$ have been searched by \citet{almeida81}
without success. Optical detections of OH$^+$ are reported by \citet{krelowski10b}.
Using the relations of \citet{mattila86}, the observed columns of
OH$^+$ and CH are used to  infer a fractional abundance of $N(\mathrm{OH}^+)/N(\mathrm{H})
= 1.1 \times 10^{-8} N(\mathrm{OH}^+)/N(\mathrm{CH}) = 9 \times 10^{-10}$, where $N(\mathrm{H}) = 2 N(\mathrm{H}_2)$.
The fractional abundance of OH$^+$ of about $10^{-9}$\ cm$^{-2}$ is almost
two orders of magnitude lower than the abundances inferred toward the massive star forming region
W31C \citep{gerin10}.  The absence of velocity shifts between OH$^+$
and the neutral interstellar molecules (cf. Table~\ref{ismolecules})
argues against a production of OH$^+$ in shocks, as proposed
by \citet{almeida90}. Interstellar shocks have been ruled out as
the general production site of CH$^+$ \citep{gredel93}.

{\bf NH.} Interstellar NH has been discovered by \citet{meyer91}.
From their measurements of the $R_1(0)\ (\lambda_\mathrm{air} =
3358.0525\ \AA$) and $^RQ_{21}(0)$ ($\lambda_\mathrm{air} =
3353.9235\ \AA$) lines of the (0,0) band of the $\mathrm A^3\Pi -
X^3\Sigma$ system, the authors inferred column densities of $1
\times 10^{12}$ cm$^{-2}$ and $2.7 \times 10^{12}$ cm$^{-2}$
toward $\zeta$ Per and HD 27778, respectively. Both lines are
detected in our spectrum of HD 169454 and are marginally present
toward CPD $-32^\circ1734$. We use oscillator strengths of
$f=0.0041$ for the $R_1(0)$ and $f=0.0024$ for the $^RQ_{21}(0)$
line \citep{meyer91} to infer a column density of $N$(NH) = $7.6
\times 10^{12}$\ cm$^{-2}$ toward HD 169454. The column density is
 in good agreement with the value of $6.7 \times 10^{12}$\ cm$^{-2}$ given recently by
\citet{weselak09a} in their comprehensive analysis of interstellar NH.
An NH column density of
$N$(NH) $\approx 2.4 \times 10^{12}$\ cm$^{-2}$ is inferred toward CPD $-32^\circ 1734$. Toward both lines
of sight, the radial velocities  of the NH absorption lines agree with those
of CH, CH$^+$, and CN.

{\bf CN.}
The $R(1)\ (\lambda_\mathrm{air} = 3579.453\ \AA$), $R(0)\ (\lambda_\mathrm{air} = 3579.963\ \AA$), and
$P(1)\ (\lambda_\mathrm{air} = 3580.937\ \AA$)  lines of the (1,0) band of the $\mathrm{B^2\Sigma - X^2\Sigma}$
system of CN are detected toward HD 169454, and the R(0) line is detected toward BD $-14^\circ$5037,
CPD $-32^\circ$1734, and CPD $\-33^\circ$1768.
We use the molecular parameters given by \citet{meyer89} to infer interstellar CN abundances of
$N(J=0) = 2.4 \times 10^{13}$ cm$^{-2}$ and $N(J=1) = 1.2 \times 10^{13}$\ cm$^{-2}$
toward HD 169454, in the limit of Doppler-$b$ values of $b \rightarrow \infty$.
The ratio in the population densities corresponds to an excitation temperature of $T_\mathrm{ex} = 2.9$ K.
The value of $N(J=0)$ inferred here is consistent with the
value of $N(J=0) = 3 \times 10^{13}$\ cm$^{-2}$
obtained for a Dopper-$b$ value of $b_\mathrm{vr} = 0.45$ km s$^{-1}$
\citep{gredel91}, where $b_\mathrm{vr}$ was obtained
from an analysis of CN absorption lines arising in the violet and red systems.
The CN column densities in $J=0$ toward BD $-14^\circ5037$, CPD $-32^\circ1734$, and CPD $-33^\circ1768$
(cf. Table~\ref{ismolecules}) are consistent with the measurements of \citet{gredel91}
and \citet{gredel02} toward those lines of sight.

{\bf CH.}
We report the detection of the
$R_2(1)\ (\lambda_\mathrm{air} = 3137.576\ \AA)$
and $^PQ_{12}(1)\ (\lambda_\mathrm{air} = 3145.996\ \AA$) lines and of the
$[Q_2(1) + ^QR_{12}(1)]\ (\lambda_\mathrm{air} = 3143.183\ \AA$) unresolved line blend of
the (0,0) band of the $\mathrm C^2\Sigma^+ - X^2\Pi$ system of CH. In addition, the
$[^QR_{12}(1) + Q_{12}(1)]\ (\lambda_\mathrm{air} = 3633.29\ \AA$) line blend and the
$^PQ_{12}(1)\ (\lambda_\mathrm{air} = 3636.27\ \AA$) line of the
(1,0) band of the $\mathrm B^2\Sigma^- - X^2\Pi$ system are detected.
The measured equivalent widths of the lines arising in the C-X system are converted into column densities
using the molecular parameters and the method of \citet{lien84}.
For the oscillator strength of the (1,0) band of the B-X system, we scale the value of the (0,0) band of
Lien (1984)  using the ratio of 7.8 of the band oscillator strengths of the (0,0) and
(1,0) bands as inferred from the lifetime measurements of \citet{brooks74}.
The observed CH absorption lines and line blends arise
from the lower or upper component of the $J = 1/2$ $\Lambda$-doublet.
In Table~\ref{ismolecules}, we list total CH column densities $N_\mathrm{tot}$, which are obtained under the assumption
that $N_\mathrm{tot} = 2N_\mathrm{l}; 2N_\mathrm{u}$,
where $N_\mathrm{l}$ and $N_\mathrm{u}$ are the column densities in the lower and upper $\Lambda$
components, respectively. For the purpose of the present
consistency check, we assume that all lines are optically thin, and ignore a detailed curve of growth analysis.

{\bf CH$^+$.}
The $R(0)$ line ($\lambda_\mathrm{air} = 3745.307\ \AA$) of the (2,0) band of the $\mathrm{A^1\Pi - X^1\Sigma^+}$ system of CH$^+$
is detected toward all five lines of sight studied here, and the
$R(0)$ line of the (3,0) band ($\lambda_\mathrm{air} = 3579.021\ \AA$) is detected toward
HD 169454, BD $-14^\circ$5037, CPD $-32^\circ1734$, and CPD $-33^\circ1768$.
We use the molecular parameters given by \citet{weselak09b}
to infer CH$^+$ column densities ranging from
$N$(CH$^+$) = $1.6 \times 10^{13}$ cm$^{-2}$ toward
HD 169454 to
$N$(CH$^+$) $\approx 4 \times 10^{13}$ cm$^{-2}$ toward BD$-14^\circ5037$, CPD$-32^\circ1734$,
 and CPD$-33^\circ1768$.

%
%
%
%
\begin{landscape}
\begin{table}
\caption{Summary of molecular absorption line measurements.  }
\label{ismolecules}
\centering
\begin{tabular}{llrrrlrrrlrrrlrrrlrrr}
\hline\hline
& \multicolumn{4}{l}{HD 169454} & \multicolumn{4}{l}{BD -14$^\circ$5037} & \multicolumn{4}{l}{CPD -33$^\circ$1768}
& \multicolumn{4}{l}{CPD -32$^\circ$1734} \\
& $\lambda_\mathrm{hel}^a$ & $W_\lambda^a$ & $V_\mathrm{hel}^1$ &$N^a$
& $\lambda_\mathrm{hel}^a$ & $W_\lambda^a$ & $V_\mathrm{hel}^1$ &$N^a$
& $\lambda_\mathrm{hel}^a$ & $W_\lambda^a$ & $V_\mathrm{hel}^1$ &$N^a$
& $\lambda_\mathrm{hel}^a$ & $W_\lambda^a$ & $V_\mathrm{hel}^1$ &$N^a$\\

OH$^+$&\ldots  & \ldots & \ldots & \ldots & \ldots & \ldots & \ldots & \ldots& \ldots&\ldots & \ldots &\ldots & 3584.193 & 4 (1)& +35.4 & 11 (3)& \\

NH    & 3353.825 & 1.8 (1)& --8.9 & 7.5 (4)  & \ldots  & \ldots&\ldots& & \ldots   & \ldots &\ldots& & 3354.3: & $\le$ 1 & \ldots & $\le$ 4 & \\
      & 3357.952 & 3.1 (1)& --9.0 & 7.6 (2) & \ldots  & \ldots&\ldots& &  \ldots   & \ldots &\ldots& & 3358.5: & $\le$ 1.0 & \ldots & $\le$ 2.4 & \\

CH    & 3137.461 &3.6 (1.0) & --11.0& 39 (11)& \ldots & \ldots&\ldots& & \ldots   & \ldots &\ldots&\ldots& 3137.954 & 7.5(1) & +36.1 &82 (11)   \\
      & 3143.069 &10.0 (2.5)& --10.8& 36 (9)& 3143.100 & 8.7 (1.5)& --7.9& 32 (5)& 3143.561 & 19(3) & +36.0&69 (11)  & 3143.554 & 18 (3)  &+35.3 & 65 (11) \\
      & \ldots   &\ldots   &\ldots &\ldots& 3143.216 &15 (1.5)& +3.2 &54 (5) & \ldots   & \ldots & \ldots & \\
      & 3145.891 &11.0 (2.5)& --10.0& 58 (13)  &          &          &\ldots& &  3146.387 & 20(3) &+37.2&106 (16)& 3146.378 & 13 (3)  & +36.3 & 69 (16) \\
      & 3633.178 & 1.1 (0.5)& --9.2& 47 (20)   & 3633.206 & 1.5 (0.5)&--6.9 &64 (20)& \ldots   & \ldots &\ldots&\ldots& 3633.738 & 2.4 (1)& +36.9 & 100(40)& \\
      & \ldots   & \ldots   &\ldots&      & 3633.36:   &$\le$ 1.5& \ldots  &$\le$ 65 & \ldots   & \ldots & \\
      & 3636.126 & 1.0 (0.5)&--11.8& 63 (32)  & 3636.144 & 1.0 (0.5)&--10.4&63 (32) & \ldots & \ldots &\ldots&\ldots& 3636.684 & 2.0 (1)  & +34.1 & 130 (63)& \\
      & \ldots   & \ldots   &\ldots&\ldots& 3636.271 & 1.5 (0.5)&+0.1  & 95 (32)   \\

CH$^+$& \ldots   & \ldots   &\ldots&  & 3579.041 & 3.6 (1) & +1.7 & 42 (11) & 3579.473 &2(0.5)&+37.8 & 24(6) & 3579.441  & 3.1 (1) & +35.1 & 36 (11) \\
      & \ldots   & \ldots   &\ldots&  & 3579.111 &0.8 (0.5)& +7.5 & 9 (6) \\
      & 3745.188 & 3.3 (0.5)&--9.5  & 16 (3)  & 3745.366 & 8.7 (1)& +4.6 & 41 (5) & 3745.767 &8 (2)& +36.7& 38 (10) & 3745.760& 8 (1)&+36.1  & 38 (5) \\

CN& 3579.344 & 2.7 (1.0)& --9.1 &12 (4) \\
& 3579.855 & 8.1 (1.5)& --9.0 & 24 (4) & 3579.887 & 4.2 (1) & --6.4 & 12 (3)& 3580.40 &3 (1) &+36.5 & 9 (3) & 3580.411 & 1.1 (0.5)& +37.5 & 3.2 (1.5) \\
& \ldots   & \ldots   &\ldots&\ldots& 3580.008 & 2.6 (1) & +3.8 & 7.6 (3) \\
& 3580.831 & 1.3 (0.5)& --8.9 & 11.5 (4) \\

\hline
\multicolumn{10}{l}{$^a$ $\lambda$ in $\AA$; $W_\lambda$ in m$\AA$;
$V_\mathrm{hel}$ in km s$^{-1}$; $N$ in $10^{12}$cm$^{-2}$}. \\
\end{tabular}
\end{table}
\end{landscape}

\section{Discussion}
\label{discussion}

\subsection{The long and eventful history of DIB candidates}

Over the last decades, a large variety of potential DIB carriers has been
proposed, either because laboratory spectra revealed coincidences with
absorption features in astronomical spectra or
because theoretical calculations predicted molecular transition
wavelengths that matched those of observed DIBs.
For instance, \citet{tulej98} proposed  C$_7^-$ based on a few
coincidences between bound-bound electronic transitions of C$_7^-$ and
five DIB absorption wavelengths. The proposition is no longer generally accepted
\citep{mccall01}, as the exact wavelengths and absorption profiles
of the C$_7^-$ lines do not match the interstellar profiles closely enough.
A number of close agreements in the absorption wavelengths
of neutral carbon chain molecules and their cations with DIB absorptions
were found by \citet{motylewski00}, yet in all cases small but significant
shifts between the laboratory and the interstellar absorption features remained.
These shifts could not be explained by radial velocity shifts of the interstellar
material nor could they be traced to calibration issues in the astronomical data.
The HC$_4$H$^+$ diacetylene cation has recently been purported as a DIB carrier
\citep{krelowski10a}. The authors averaged a large number of stellar spectra
of different stars in order to improve the S/N-ratio. The average spectrum
reveals a faint absorption feature near
5069 $\AA$ which \citet{krelowski10a} assign to a previously
undetected diffuse interstellar band. The gas-phase spectra obtained by
\citet{motylewski00} shows a prominent absorption band of HC$_4$H$^+$ at 5069 $\AA$,
which matches exactly the position of the 5069 $\AA$ feature.
However, the HC$_4$H$^+$ band profile obtained from the laboratory data deviates
significantly from the 5069 $\AA$ feature, and it is difficult to see
how a convolved HC$_4$H$^+$ band profile, assuming typical  Doppler-$b$
broadening of the rotational lines and macroscopic turbulent gas motions in translucent clouds,
can be brought into agreement with the profile of the 5069 $\AA$ absorption feature
reported by \citet{krelowski10a}.

In a similar fashion, the recent evidence provided by \citet{maier10} that
$l$-C$_3$H$_2$ is responsible
for the 5450 $\AA$ and 4881 $\AA$ DIBs requires some scrutiny.
While the authors convincingly assign two absorption features
near 4887 $\AA$ and 5450 $\AA$ seen in their  laboratory CRDS spectra
to the $2^1_0$ and $2^2_0$ transitions in the
$B^1\mathrm{B}_1 \leftarrow X^1\mathrm{A}_1$ system of $l$-C$_3$H$_2$,
the agreement with interstellar DIBs is good for the 5450 $\AA$  line
only. The 4887 $\AA$ transition of $l$-C$_3$H$_2$ does not match the exact
shape and position of the 4881 $\AA$ DIB, unless it is assumed that several DIB
carriers contribute to this feature.  More critically, Ne-matrix spectra
also reveal an $l$-C$_3$H$_2$ absorption feature which arises from the
$2^1_04^1_0$ transition, which is about a factor of 2 -- 3 fainter than
the 4887 $\AA$ and 5450 $\AA$ transitions of $l$-C$_3$H$_2$.
The strength of the $2^1_04^1_0$ transition is below the detection
limit of the CRDS experiments presented by \citet{maier10} and an accurate gas phase
wavelength is not available, yet its position  is predicted to be near 5173 $\AA$.
\citet{maier10} did search their spectra for
an interstellar absorption band near 5173 $\AA$, yet they note that the
presence of various stellar absorption lines hampered their effort.
The high S/N spectrum presented by \citet{hobbs09} toward HD~183143 does show
broad and prominent DIB absorptions near 4881 $\AA$ and 5450 $\AA$. The latter spectrum
is relatively clean in the 5165 -- 5180 $\AA$ region, yet it does not show
the expected signature from  the $2^1_04^1_0$ transition of $l$-C$_3$H$_2$ in this
wavelength interval.

Other molecular cations and anions have served as prominent
candidates of DIB carriers in the past. Matrix isolation
experiments of C$_{60}^+$ revealed absorption features near the
9577 $\AA$ and 9632 $\AA$ DIBs \citep{foing97}, however a
confirmation of the C$_{60}^+$ gas phase wavelengths based on CRDS
spectra is not yet available. That work  supported the general
proposition that PAH cations are responsible for DIB absorptions
\citep{Salama96, Salama99}. Jet-cooled absorption spectra measured
in the laboratory \citep{Romanini99, Biennier03, Biennier04,
Sukhorukov04, Tan06} were subsequently obtained, and interstellar
spectra obtained toward the Perseus molecular cloud complex did
reveal weak features that coincide with absorption bands of the
naphthalene and anthracene cations \citep{IglesiasGroth08,
IglesiasGroth10}. {In the case of the anthracene
cation, the CRDS spectrum of \citet{Sukhorukov04} reveals an
absorption at 7085.7 $\AA$ which has a FWHM of 47 $\AA$.} While
the spectrum of \citet{IglesiasGroth10} may show an interstellar
absorption band near that wavelength, the width of the purported
interstellar absorption feature is significantly smaller than 47
$\AA$.

Potential DIB carriers have also been proposed from theory. With
rotational contour fitting, \citet{kerr96} were able to reproduce
the structure seen in the 6614 $\AA$ and 5797 $\AA$ DIBs assuming
that they are due to planar oblate symmetric top molecules, such
as large carbon ring molecules with 14 -- 30 atoms, rather than
PAHs. An excellent agreement was found between the calculated
absorption wavelength of the $0^0_0$ band of the $^1B_1 - X^1A'$
electronic system of  CH$_2$CN$^-$ and the 8037 $\AA$ DIB
\citep{cordiner07}. However, the calculated absorption wavelength
needs to be confirmed by gas phase laboratory experiments. It is
also noted that the match between the calculated CH$_2$CN$^-$
absorption and the 8037 $\AA$ DIB is good for an assumed
rotational excitation temperature of 2.7~K and a Doppler-$b$ value
of 16 - 33 km s$^{-1}$. At smaller $b$-values, the rotational
structure of CH$_2$CN$^-$ should be resolved in the astronomical
spectra, in disagreement with the observed broad and diffuse
absorption near 8037 $\AA$. It is also noted that at a nominal
ortho to para ratio of 3:1, the calculated spectrum of
CH$_2$CN$^-$ does not match the observed spectra. Similarly, a
recent claim that proflavine and its anion, cation, and di-cation
are responsible for the 4066 $\AA$, 4363 $\AA$, 4175 $\AA$, and
5259 $\AA$ DIBs \citep{bonaca10} requires verification. The
proposition is based on computed optical absorption wavelengths
using pseudo-potential density functional theory methods. The
matches in wavelengths are marginal at most, and need to be
confirmed by laboratory experiments. {In any case,
as pointed out at the end of Sect.~\ref{lab-sigma}, extreme care
must be taken if computational results on electronic transitions
are used for the interpretation of astronomical  observations, due
to the large uncertainties still being involved in present-day
calculations.}

\subsection{Fractional PAH abundances in translucent molecular clouds}
\label{pahabundance}

PAHs are believed to be formed in the outflows of carbon-rich AGB stars from where they are deposited
into the interstellar medium. The fractional PAH abundance is estimated to be of the order of
$f$(PAH) = $N$(PAH)/$N$(H) = $10^{-7}$ \citep{habart04}, assuming some 100 carbon atoms
per PAH molecule. In the diffuse interstellar material,
at densities of $n_\mathrm{H}$ = 100 cm$^{-3}$, a temperature of
$T = 100$~K, and fractional electron abundances
of $n_e = 10^{-4}$, small PAHs such as anthracene, phenanthrene, and pyrene, are rapidly
destroyed by the ambient UV radiation \citep{page03}. The photoionization of PAHs in diffuse
clouds is a major heating source \citep{lepp88c} and supersedes the grain heating for PAH
abundances greater than $2 \times 10^{-7}$.
As the visual extinction increases, the abundance of PAHs increases rapidly, as shown by
\citet{lepp88b}.  The authors postulated a fractional PAH abundance  of $10^{-7}$
in an attempt to resolve density discrepancies which persisted for translucent molecular cloud models.
The transition between neutral PAHs and their anions and cations is governed by the balance between
photoionization and recombination as well as electron attachment and photo-detachment.
At visual extinctions of $A_\mathrm{V}$ = 1 mag (which corresponds to a reddening of $E_\mathrm{B-V}$ = 0.3 mag),
the fractional abundances of neutral PAH and their anions are comparable and exceed the fractional abundance
of the positively charged PAHs by about two orders of magnitude.
As the visual extinction increases further, PAHs form a drain to the free electrons for PAH abundances greater than
$10^{-8}$, resulting in a substantial reduction in the density of electrons \citep{lepp88a}.
As a result, the molecular ions undergo nondestructive neutralization reactions with the negatively
charged PAHs, which leads to a significant increase of the fractional abundances of
the carbon-bearing molecules.

Based on the infrared emission bands, \citet{Allamandola85}
evaluated the fractional abundance of interstellar PAHs to be $2
\times 10^{-7}$. Later, \citet{Allamandola89} distinguished
between small and large PAHs, which were defined as containing 25
and 300-400 C atoms, respectively. A minimal fractional abundance
of $3 \times 10^{-8}$ was obtained for the small PAHs while a
value of $8 \times 10^{-9}$ was derived for the large PAHs
\citep{Allamandola89}, corresponding to fractions of a 1 -- 5\%
and 0.5\% of elemental carbon locked in small and large PAHs,
respectively. Considering a single class of PAHs with a typical
size of 50 C atoms, an abundance of about $3 \times 10^{-7}$ was
derived by \citet{Tielens08}. Of interest to our study is the
minimal interstellar fractional abundance of the small PAHs. With
the exception of hexabenzocoronene, the molecules we have studied
can be considered to belong to the class of small PAHs. The upper
limits we have derived for the fractional abundances of
anthracene, pyrene, and 2,3-benzofluorene are about two to three
orders of magnitude lower than the estimates presented by
\citet{Tielens08}. The fractional abundance of phenanthrene is not
well constrained from our observations, mainly due to the fact
that the phenanthrene band studied here has a small oscillator
strength. We would thus conclude that small and neutral PAHs must
occur in a large variety (several 100 species) in translucent
clouds, if it is assumed a fraction of several percent of the
elemental carbon is locked up in small PAHs. Alternatively, small
and neutral PAHs may not be very abundant in translucent material,
either because small PAHs carry positive or negative charge or
because significantly less elemental carbon is locked up in these
molecules. 

In their investigation of the ultraviolet interstellar extinction
curve using the HST/STIS spectrograph, \citet{clayton2003} 
did not detect absorption bands in the 1150 $\AA$ - 3180 $\AA$ 
wavelength region that could be associated with PAHs. The authors
conlcuded that either PAHs are destroyed by the interstellar
radiation field, or that they occur in such a large variety that
individual absorptions are too weak to be detected.

Concerning the larger PAHs, an upper limit in the column density
of hexabenzocoronene of $4 \times 10^{12}$\ cm$^{-2}$ in translucent
molecular material has recently been estimated by
\citet{Kokkin2008} from spectra covering the vibronic
hexabenzocoronene absorptions in the $\alpha$-band near 4262
$\AA$.  The $\alpha$-band is, however, not well suited for the
search of interstellar hexabenzocoronene, as it is significantly
weaker than the $\beta$-band near 3344 $\AA$ \citep{Rouille2009}.
Unfortunately, we are not able to constrain the molecular
abundance of hexabenzocoronene from our observations, as we were
unable to derive absorption cross sections of the $\beta$-band
(cf. Sect.~\ref{lab-sigma} and Sect.~\ref{results}). We
nevertheless propose that searches for interstellar
hexabenzocoronene be conducted near 3344 $\AA$.

\subsection{Molecular anions in translucent material}

Carbon-chain molecular anions have recently been established as a major constituent of dense material.
The detection of C$_6$H$^-$  in the TMC-1 molecular cloud by
\citet{mccarthy06} marks the discovery of molecular anions in the interstellar medium. That work
was followed by the observation of several other anions in interstellar and circumstellar material, e. g.
C$_4$H$^-$ \citep{cernicharo07, agundez08}, C$_8$H$^-$ \citep{remijan07}, $\mathrm{C}_3
\mathrm{N}^-$ \citep{thaddeus08}, and CN$^-$ \citep{agundez10}.
The effects of molecular anions on the chemistry of dark clouds have been rediscussed
by \citet{millar07} and \citet{walsh09}.  The models set up for dark clouds
produce large amounts of molecular anions at molecular cloud ages below $10^5$ yr. The anion abundances
are proportional to the electron abundance and are characterized by a sharp drop by several orders of
magnitude as the clouds reach ages of $10^6$ yr or above. The anion to neutral
ratio is typically of the order of a few percent.

\subsection{Individual lines of sight}

{\bf HD 169454\ and\ BD -14$^\circ$5037.}
Various studies of the lines of sight toward HD 169454
and BD $-14^\circ$5037 are available in the literature. The C$_2$ observations of \citet{gredel86}
revealed the presence of very cold gas toward HD 169454, with gaskinetic temperatures of $T_\mathrm {kin}$ = 15~K
and gas densities of $n$ = 400 cm$^{-3}$. The C$_2$ absorption occurs at heliocentric velocities of
 $V_\mathrm{hel}$ = --8.5 km s$^{-1}$.
Toward BD $-14^\circ 5037$, the C$_2$ absorption occurs in two velocity components at
$V_\mathrm{hel}$ = --6.0 km s$^{-1}$ and $V_\mathrm{hel}$ = + 4.0 km s$^{-1}$, where the gas in the two
velocity components is characterized by
temperatures and densities of $T_\mathrm {kin}$ = 30~K and $n$ = 400 cm$^{-3}$ and
$T_\mathrm {kin}$ = 50~K and $n$ = 250 cm$^{-3}$, respectively. The CH and CN absorption
lines are resolved into two velocity components at similar heliocentric velocities
(cf. Table~\ref{ismolecules}). The CH$^+$ absorption profiles are
unresolved and appear near heliocentric velocities which are consistent with those of C$_2$, CH, and CN.
Very small Doppler-$b$ values of $b = 0.4$ km s$^{-1}$ toward both lines of sight
were inferred by \citet{gredel91} from
a comparison of CN column densities derived from the (0,0) bands of the CN red and violet systems.
A curve of growth analysis
using $b = 0.4$ km s$^{-1}$ resulted in total CN column densities of $N$(CN) = $4.2 \times 10^{13}$ cm$^{-2}$
and $N$(CN) = $1.6 \times 10^{13}$ cm$^{-2}$ toward HD 169454 and BD$-14^\circ5037$, respectively, and
in very good agreement with the present results.
The present data are obtained from absorption lines in the (1,0) band, which are about a factor of 10
weaker than the absorption lines in the (0,0) band. The excellent agreement between both sets of
measurements confirms the very small Doppler-$b$ values toward both lines of sight.
Using the small Doppler-$b$ values, \citet{gredel93} obtained CH and CH$^+$ column densities of
$N(\mathrm{CH}) = 4.4 \times 10^{13}$\ cm$^{-2}$ and $N(\mathrm{CH}^+) = 1.8 \times 10^{13}$\ cm$^{-2}$
toward HD 169454, and
$N(\mathrm{CH}) = 7.3 \times 10^{13}$\ cm$^{-2}$ and $N(\mathrm{CH}^+) = 6.4 \times 10^{13}$\ cm$^{-2}$
toward BD $-14^\circ5037$, again in excellent agreement with the present results.

{\bf CPD --32$^\circ$1734 and CPD --33$^\circ$1768.} Very large
column densities of C$_2$, CH, CH$^+$, and CN have previously been
measured toward the heavily reddened supergiants CPD $-32^\circ
1734$  and CPD $-33^\circ 1768$ in the NGC 2439 association
(Gredel et al. 2002, and references therein). From observations of
C$_2$, gaskinetic temperatures of $T_\mathrm{kin} = 85$ K,
densities of $n > 1000$~cm$^{-3}$, and column densities of
$N(\mathrm{C}_2) \approx 10^{14}$\ cm$^{-2}$ were inferred toward
both lines of sight. The CH and CH$^+$ column densities reach
values of $N$(CH$^+$) = $4.4 \times 10^{13}$\ cm$^{-2}$  and
$N$(CH) = $9 \times 10^{13}$\ cm$^{-2}$ toward CPD $-32^\circ
1734$ and $N$(CH$^+$) = $6.0 \times 10^{13}$\ cm$^{-2}$  and
$N$(CH) = $12 \times 10^{13}$\ cm$^{-2}$ toward CPD $-33^\circ
1768$. The column densities and heliocentric velocities obtained
here are in excellent agreement with the previously published
results.

{\bf HD 183143.}
A comprehensive optical study of the diffuse interstellar bands toward HD 183143
has recently been introduced by \citet{hobbs09}.  The authors present a catalog of 414 DIBs
from spectra obtained in the 3900 - 8100 $\AA$ wavelength region with a signal-to-noise
ratio of S/N $\approx 1000$ and
a spectral resolution of $R$ = 38\,000. Their study forms part of a new survey of DIBs
toward 30 OB stars carried out at the Apache Point Observatory at a very high signal-to-noise ratio.
Molecular carbon was not detected toward HD 183143 \citep{gredel99}, which indicates that the
line of sight is characterized by low-density material.

\section{Summary}
We have presented high signal-to-noise ($S/N >$ 100) absorption line spectra toward heavily reddened
early type supergiants in the 3050 -- 3850 $\AA$ wavelength region. The spectra are compared with
laboratory gas-phase spectra of anthracene, phenanthrene, pyrene, 2,3-benzofluorene, and benzo[ghi]perylene.
We have developed methods to infer absolute absorption cross sections from the laboratory
spectra, which we use to derive upper limits in the column densities and the abundances
of these neutral PAHs. Our methods lead to reasonable values for the absorption cross sections,
if compared with cross sections derived from {\it ab initio} methods. The fractional abundances for the
PAHs are below a few times $10^{-10}$ for anthracene, pyrene, and 2,3-benzofluorene. Fractional abundances
of benzo[ghi]perylene below a few times $10^{-9}$ and of phenanthrene below a few times $10^{-8}$ are
derived. Thus, a large variety of small and neutral PAHs such as anthracene and pyrene is required to
explain a canonical value of $10^{-7}$ for the total fractional PAH abundance in the interstellar medium.
Alternatively, it may be concluded that small and neutral PAHs are suppressed in translucent molecular
clouds, either because the small PAHs
carry predominantly positive or negative charge or because they don't
survive the prevailing physical conditions in such clouds.

\begin{acknowledgements}
Discussions with J. Kre{\l}owski and with S. Federman are kindly acknowledged.
F.H., G.R., and Y.C. acknowledge the financial support of the Deutsche
Forschungsgemeinschaft (DFG).
\end{acknowledgements}

\end{document}